\documentclass[fleqn,usenatbib]{mnras}

\usepackage{newtxtext,newtxmath}

\usepackage[T1]{fontenc}

\DeclareRobustCommand{\VAN}[3]{#2}
\let\VANthebibliography\thebibliography
\def\thebibliography{\DeclareRobustCommand{\VAN}[3]{##3}\VANthebibliography}

\usepackage{graphicx}	
\usepackage{amsmath}
\usepackage{comment}
\usepackage{subfigure}

\newcommand{\Bk}{\mathrm{Bk}}

\title[Mode Conversion in a Magnetar Atmosphere]{X-ray Polarisation in Magnetar Atmospheres - Effects of Mode Conversion}

\author[R. M. E. Kelly et al.]{
Ruth M. E. Kelly,$^{1}$\thanks{E-mail: ruth.kelly.22@ucl.ac.uk}
Silvia Zane,$^{1}$
Roberto Turolla,$^{2,\,1}$
Roberto Taverna$^{2}$
\\
$^{1}$Mullard Space Science Laboratory, University College London, Holmbury St Mary, Dorking, Surrey RH5 6NT\\
$^{2}$Universit\`a di Padova, Dipartimento di Fisica e Astronomia, via Marzolo 8, I-35131 Padova, Italy\\
}

\date{Accepted XXX. Received YYY; in original form ZZZ}

\pubyear{2024}

\begin{document}
\label{firstpage}
\pagerange{\pageref{firstpage}--\pageref{lastpage}}
\maketitle

\begin{abstract}
Magnetars, the most strongly magnetised neutron stars, are among the most promising targets for X-ray polarimetry. The Imaging X-ray Polarimetry Explorer (IXPE), the first satellite devoted to exploring the sky in polarised X-rays, has observed four magnetars to date. 
A proper interpretation of IXPE results requires the development of new atmospheric models that can take into proper account the effects of the magnetised vacuum on par with those of the plasma.
Here we investigate the effects of mode conversion at the vacuum resonance on the polarisation properties of magnetar emission by 
computing plane-parallel atmospheric models under varying conditions of magnetic field strength/orientation, effective temperature and allowing for either complete or partial adiabatic mode conversion. Complete mode conversion results in a switch of the dominant polarisation mode, from the extraordinary (X) to the ordinary (O) one, below an energy that decreases with increasing magnetic field strength, occurring at $\approx 0.5\, \mathrm{keV}$ for a magnetic field strength of $B=10^{14}\, \mathrm{G}$. Partial adiabatic mode conversion results in a reduced polarisation degree when compared with a standard plasma atmosphere. No dominant mode switch occurs for $B=10^{14}\, \mathrm{G}$ while there are two switches for lower fields of $B=3\times10^{13}\, \mathrm{G}$.
Finally, by incorporating our models in a ray-tracing code, we computed the expected polarisation signal at infinity for different emitting regions on the star surface and for different viewing geometries. The observability of QED signatures with IXPE and with future soft X-ray polarimeters as REDSoX is discussed.

\end{abstract}




\section{Introduction}

Magnetars are (isolated) neutron stars (NSs) characterised by their super strong dipole magnetic fields of $B\approx 10^{14}$--$10^{15}$ G \citep{thompson_neutron_1993}. Powered by their own magnetic energy, magnetars produce bursts of hard X-/soft $\gamma$-rays as well as persistent emission which spans from radio to soft $\gamma$-rays. Observationally identified as the Soft $\gamma$-repeaters (SGRs) and the Anomalous X-ray pulsars (AXPs), they shine in  X-rays with luminosities $L \approx 10^{31}-10^{36}$ erg s$^{-1}$, exhibit spin periods $P \approx 1-12$ s{\footnote{There are indications that the true period of the transient magnetar 3XMM J185246.6+003317 is actually $\sim 23\, \mathrm s$ \citep{hambaryan_3xmm_2015}, twice that originally reported by \cite{rea_3xmm_2014}.} and period derivatives $\dot{P} \approx 10^{-13}-10^{-10}$ s s$^{-1}$ \cite[see e.g.][for reviews]{turolla_magnetars_2015, kaspi_magnetars_2017}. 

The soft X-ray spectrum ($\sim1$--$10$ keV) is typically modelled in terms of  two thermal (blackbody) or one thermal and one non-thermal (power-law) components.
Thermal photons are believed to come from the stellar surface, covered by either a magnetic condensate or a gaseous atmosphere (see e.g. \citealt{taverna_x-ray_2020, caiazzo_probing_2022}), while the non-thermal component, when present, is thought to arise from magnetospheric effects, such as resonant cyclotron scattering (RCS) onto mildly relativistic charges \cite[see e.g.][]{thompson_electrodynamics_2002, nobili_x-ray_2008, turolla_magnetars_2015}.

In the presence of a strong magnetic field, electromagnetic waves travel through a plasma in two linear polarisation modes \citep{gnedin_effect_1978}.
These ``normal modes'' are referred to as the Ordinary (O) and Extraordinary (X) modes and have the polarisation electric vector parallel or perpendicular to the plane of the propagation direction and magnetic field, respectively.
The two modes have very different opacities. The X-mode photons propagate with a reduced refractive index because the wave electric field is perpendicular to the local magnetic field. This results in a reduction of the cross sections so that they do not interact with the electrons in the plasma as effectively as O-mode photons, the opacities of which are quite unchanged with respect to the non-magnetic case.
Emission from a magnetar is then expected to be polarised, with the degree of polarisation depending on the physical state of the outer layers of the star and on the processes occurring in the magnetosphere.

The launch of the NASA-ASI Imaging X-ray Polarimetry Explorer (IXPE; \citealt{weisskopf_imaging_2022}) in late 2021 allowed us to systematically study the polarisation of X-ray sources for the first time. To date, IXPE has observed four magnetar sources: the AXPs 4U 0142+61 \citep{taverna_polarized_2022}, 1RXS J170849.0-400910 \cite[hereafter 1RXS J1708 for short,][]{zane_strong_2023}, 1E 2259+586 \citep{heyl_detection_2024} and
SGR 1806-20 \citep{turolla_ixpe_2023}.
 
With an inferred dipole magnetic field of $B \approx 1.3\times10^{14}$ G at the equator, 4U 0142+61 was found to have an X-ray polarisation which varies considerably throughout the $2$--$8$ keV IXPE energy band \citep{taverna_polarized_2022}.
At lower energies, $2$--$4$ keV, the magnetar polarisation degree is $\approx 15\%$ and it increases to $\approx 35\%$ in the $5.5$--$8\,\mathrm{keV}$ range. Interestingly, at $4$--$5\,\mathrm{keV}$ there is a $90^\circ$ swing in polarisation angle, while the polarisation degree touches zero, indicating that the dominant polarisation mode is different in the high and low energy ranges.
\citet{taverna_polarized_2022} suggested that the polarisation properties of 4U 0142+61 can be explained by the reprocessing of thermal radiation from a condensed surface by RCS. The $2$--$4$ keV energy range is dominated by O-mode photons while the X-mode dominance in the $5.5$--$8$ keV range is a result of RCS occurring in the magnetosphere.

1RXS J1708, with an equatorial magnetic field of $B \sim 4$--$5\times10^{14}$ G, was also found to have a polarisation signal which varies with energy. The polarisation degree ranges from $\sim 20\%$ at $2$--$3$ keV to $\sim 80\%$ at $6$--$8$ keV. However, unlike 4U 0142+61, the polarisation angle remains constant throughout \citep{zane_strong_2023}. These authors concluded that the high polarisation degree in the $6$--$8$ keV energy range can be explained by standard atmospheric emission. The lower polarisation degree around  $2$--$3$ keV could be caused by emission from a warm condensed region on the surface. In this picture, emission across the entire energy range is believed to be dominated by the X-mode.

SGR 1806--20, which hosts the highest recorded characteristic magnetic field of $B \sim 10^{15}\, \mathrm G$, was found to have a polarisation degree of $31.6\pm10.5\%$ at $4$--$5\, \mathrm{keV}$, but only upper limits could be placed at lower and higher energies \cite[$24\%$ in the $1$--$4$  and $55\%$ in the $5$--$8\, \mathrm{keV}$ band, respectively, at 3$\sigma$ confidence level,][]{turolla_ixpe_2023}. Despite the difference in polarisation properties of the two neutron stars, \citet{turolla_ixpe_2023} demonstrated that, much like with 4U 0142+61, the emission from SGR 1806--20 is compatible with RCS of radiation from a condensed surface.

The most recently observed magnetar, 1E 2259+586, has a slightly lower inferred characteristic magnetic field of $B \approx 6\times10^{13}\, \mathrm{G}$. However, the source exhibits a phase-dependent absorption feature which, if interpreted as a proton cyclotron line, points at a much stronger field, $\approx 10^{14}-10^{15}\, \mathrm{G}$, close to the surface. The polarisation is modest ($\approx 20\%$) and changes with phase. \cite{heyl_detection_2024} proposed that a baryon-loaded magnetic loop is responsible for both the absorption feature and the larger polarisation detected at the primary minimum of the pulse, since resonant scattering in the loop favours the emergence of O-mode photons.

Recently, however, \citet{lai_ixpe_2023} argued that this interpretation is questionable, because the surface temperature and magnetic field strength of these sources are such that a phase transition to a condensed state is unlikely (at least when the value of the condensation temperature, which is poorly known, is taken at face-value). Moreover, in the case of 4U 0142+61, the proposed scenario implies that the spin axis of the star is close to being orthogonal to the direction of proper motion, while there is evidence that the spin axis and the proper motion in some neutron stars are aligned \cite[some orthogonal geometries have been observed though;][]{colpi_formation_2002, posselt_gemingas_2017, liu_radio_2023}. 
Instead, \citet{lai_ixpe_2023} proposed an alternative picture in which the features observed in the polarisation signal from 4U 0142+61 are caused by partial adiabatic mode conversion at the vacuum resonance, i.e. the conversion of a photon from one mode to another due to the concurrent contributions of plasma and quantum electrodynamics (QED) vacuum effects in the NS atmosphere \cite[see e.g.][]{adler_photon_1971, pavlov_effect_1979, ho_ii_2003}.

In this paper we study the effects of mode conversion on the emission spectrum of a strongly magnetised neutron star atmosphere, to investigate in what cases this may cause detectable features in the polarisation properties of magnetars and X-ray emitting neutron stars and, in particular, reproduce those of the sources observed with IXPE. We perform a self-consistent treatment of the radiative transfer through the atmosphere, including both free-free and scattering opacities. 
The paper is organised as follows: in section \ref{sec:vacandmode}, we lay out the theoretical basis of the mode conversion mechanism and in section \ref{sec:code} we detail the atmospheric numerical calculation used and the assumptions made. In section \ref{sec:results}, we present the numerical results and in \ref{sec:application} we apply our findings to observed sources. Discussion follows in section \ref{sec:discussion}.

\section{Theoretical Background}
\label{sec:theory}

Over the last three decades, many investigations have been devoted to modelling the transport of radiation through a magnetised neutron star atmosphere. However, despite this, the effects of QED mode conversion on the polarisation remain, as yet, not completely investigated. 

The early works by \citet{romani_model_1987}, \citet{shibanov_model_1992}, \citet{pavlov_model_1994}, \citet{pavlov_multiwavelength_1996}, and \citet{rajagopal_model_1996} self-consistently modelled fully-ionised neutron star atmospheres with low and moderate magnetic fields (up to $B \approx 10^{12}$--$10^{13}$ G), and different chemical compositions. Emerging spectra were found to be not very different from a blackbody at the star effective temperature, apart from a distinctive hardening that tends to be less prominent at larger $B$-fields. These models set the basic framework that was then used in subsequent investigations, i.e. the cold plasma and normal mode approximations \citep{ginzburg_propagation_1970}. Model atmospheres have been computed in plane parallel approximation, assuming a constant, magnetic field parallel to the slab normal and accounting for thermal bremsstrahlung and Thomson scattering as contributions to the total opacity. However, only electrons and protons were included in these models and vacuum polarisation was not accounted for.

The discovery of magnetars and other highly magnetised neutron stars prompted several efforts to extend these calculations at higher $B$-fields, and models with $B \gtrsim 10^{14}$ G have been presented by \citet{zane_magnetized_2000}, \citet{ho_atmospheres_2001} \citet{ozel_surface_2001}, \citet{lai_matter_2001}, and \citet{lloyd_model_2003}. Some of these codes also successfully accounted for different field inclinations with respect to the slab, or, as in \cite{zane_magnetized_2000} for energy deposition from accreting material. In parallel, the role of the vacuum contribution to the opacities, that becomes important above $B_{\mathrm Q}=m_ec^3/\hbar e\sim 4.4\times10^{13}\, \mathrm{G}$, has been appreciated and vacuum effects began to be included in atmosphere models, with the work of \citet{zane_magnetized_2000}, \citet{zane_proton_2001}, \citet{lai_resonant_2002}, \citet{ozel_surface_2001}, \citet{ho_iii_2003}, and \citet{lloyd_model_2003}. These works highlighted the existence of density-dependent vacuum resonances, and the challenge of computing numerical models in the presence of strongly peaked opacity coefficients. Besides, the breakdown of the normal modes approximation made necessary the introduction of an ad hoc assumption on the behaviour of the photon polarisation state near the vacuum resonances. In these seminal works only the two limiting cases of either no or complete mode conversion were considered (see \S ~\ref{sec:vacandmode} for more details). The main problem is that, while this assumption may not be too critical when one is interested in the calculation of the total spectrum, it can dramatically affect the fractional contribution of the two normal modes to the total intensity and hence the polarisation spectrum. \citet{lai_resonant_2002, ho_ii_2003}, and \citet{lai_transfer_2003} provided approximated recipes to deal with partial mode conversion but did not compute synthetic models for the emergent polarisation spectra.

Among other results, the inclusion of the vacuum contribution in the opacities motivated a reanalysis of the formation of the cyclotron line by \citet{van_adelsberg_atmosphere_2006}. Atmospheric models were improved again through the inclusion of partial ionisation \citep{ho_iii_2003, potekhin_electromagnetic_2004} and later different atmospheric compositions \citep{mori_modelling_2007}. \citet{ho_modeling_2007} investigated the emission from a thin hydrogen atmosphere above a condensed neutron star surface while cyclotron harmonics were studied by \citet{suleimanov_magnetized_2012}.

The only attempts to overcome the ad hoc assumption of no or complete mode conversion through numerical calculations have been published only very recently. \citet{gonzalez-caniulef_atmosphere_2019} included the effect of particle bombardment and returned to the question of how mode conversion impacts the spectra, investigating partial mode conversion as described by \citet{pavlov_effect_1979}. Although this work deals only with the frequency-integrated case, the analysis of the temperature profile hints at the conclusion that the polarisation degree should be substantially reduced with respect to the case in which no or complete mode conversion is assumed. More recently, \citet{lai_ixpe_2023} addressed partial adiabatic mode conversion through a magnetised atmosphere under a number of simplifying assumptions (e.g. considering only two outgoing photon rays), but without a full numerical modelling.

\subsection{Vacuum Resonance and Mode Conversion} \label{sec:vacandmode}

The presence of magnetic fields with strengths around and above the quantum critical field $B_{\mathrm Q}$, such as those hosted in magnetars, produces sizable effects on the propagation of radiation, both in a plasma and in vacuo. One of the most striking consequences, which has yet to be experimentally tested, is that strongly magnetised vacuum becomes birefringent (see e.g. \citealt{adler_photon_1971}). This QED phenomenon occurs when the magnetic field is strong enough to significantly affect the virtual electron-positron pairs that continuously create and annihilate in the vacuum around the star, causing its dielectric and magnetic permeability tensors to deviate from unity.
As a consequence, in a strongly magnetised environment both the plasma and the vacuum contribute to the dielectic and magnetic permeability tensors. At the point in the parameter space where the two contributions balance, a ``vacuum resonance'' occurs and the eigenvalues of the wave propagation equation (which are related to the oscillation direction of the electric field) become degenerate. 

In the case of a system including a plasma of free electrons and ions, as well as vacuum, the components of the wave electric vector in the plane orthogonal to the photon momentum $\boldsymbol{k}$ can be written as \citep{lai_transfer_2003,harding_physics_2006}
\begin{equation}\label{polveccomp}
\boldsymbol{E}=\frac{1}{\sqrt{1+K_\pm^2}}\left(\begin{array}{c}
iK_\pm \\ 1
\end{array}\right)\,,
\end{equation}
in a reference frame  $(\boldsymbol e_\mathrm{x},\boldsymbol e_\mathrm{y},\boldsymbol e_\mathrm{z})$, with $\boldsymbol e_\mathrm{z}$ along $\boldsymbol{k}$ and $\boldsymbol e_\mathrm{y}$ orthogonal to the $\boldsymbol{k}$-$\boldsymbol{B}$ plane. In equation (\ref{polveccomp})
\begin{equation}\label{elli} 
K_\pm = -i\frac{E_\mathrm{x}}{E_\mathrm{y}} = \beta \pm \sqrt{\beta^2 + r} \, 
\end{equation}
is the wave ellipticity referred to the basis with unit vectors  $\boldsymbol e_\pm=(\boldsymbol e_\mathrm x \pm i\boldsymbol e_\mathrm y)/\sqrt 2$, $r\simeq 1$ and the ``polarisation parameter'' $\beta$ contains all the dependence on the photon energy and direction, magnetic field strength and direction, and plasma density \cite[see Equation (18) in ][]{lai_transfer_2003}. 

In general, photons are elliptically polarised and  $\beta$ fixes the properties of the polarisation ellipse. In an ultra-magnetised environment, under most circumstances, either the vacuum or the plasma contributions to the dielectric tensor are dominant. In these cases it is typically $|\beta|\gg 1$, resulting in $|K_\pm|\gg 1$ or $|K_\pm|\ll 1$ \cite[e.g.][]{harding_physics_2006}. This translates into linearly polarised radiation, with electric field either parallel (O-mode) or perpendicular (X-mode) to the $\boldsymbol{k}$-$\boldsymbol{B}$ plane [see Equation (\ref{polveccomp})].
On the other hand, for $|\beta|\lesssim1$ the usual modal description breaks down and the polarisation state of the wave is ill defined. The points where this happens are referred to in literature as ``mode collapse points''. In particular, for $|\beta|=0$ it turns out that $|K_\pm|\simeq1$ and the wave becomes circularly polarised. 

The parameter $\beta$ is in general a complex quantity, in which $\Im({\beta})$ reflects the damping of the wave by the electrons in the plasma. When $|\beta|\gg 1$ it is also $|\Re(\beta)|\gg|\Im(\beta)|$, so that $|\beta|=|\Re(\beta)$|. This is also the case in the limit in which the damping vanishes, when $\Re(\beta)$ can be written as $\Re (\beta) = \beta_0 \beta_\mathrm{V}$ \cite[see][]{lai_transfer_2003}, where
\begin{equation}
    \beta_0 \simeq \frac{u_\mathrm{e}^{1/2}\sin^2\theta_\Bk}{2\cos\theta_\Bk}(1-u_\mathrm{i}), 
\end{equation}
\begin{equation}
    \beta_\mathrm{V} \simeq 1 + \frac{(q+m)(1-u_\mathrm{e})}{u_\mathrm{e}v_\mathrm{e}}.
\end{equation}
Here $\theta_\Bk$ is the angle between the photon direction and $\boldsymbol B$, $u_\mathrm{i} = \omega_\mathrm{Bi}/\omega$, $u_\mathrm{e} = \omega_\mathrm{Be}/\omega$ and $v_\mathrm{e} = \omega_\mathrm{pe}/\omega$, where $\omega_\mathrm{Bi}$, $\omega_\mathrm{Be}$ and $\omega_\mathrm{pe}$ are the ion, electron cyclotron and the electron plasma frequency, respectively; $q$ and $m$ are functions of the magnetic field \cite[we used the complete expressions, valid for all values of $B/B_{\mathrm{Q}}$, as given in][see also the Appendix in \citealt{potekhin_electromagnetic_2004}]{heyl_analytic_1997}.
A mode collapse point appears when $\Re(\beta)=0$, i.e. either $\beta_0=0$ or $\beta_\mathrm{V}=0$. The former condition is satisfied for $\theta_\Bk =0$, i.e. for photons propagating along the local $B$-field, or at the ion cyclotron frequency ($u_\mathrm{i}=1$). On the other hand, for frequencies well below the electron cyclotron frequency, the latter is met for $\nu_\mathrm{e}=q+m$, i.e. when the plasma and vacuum contributions balance. This is the so-called ``vacuum resonance'', which occurs when radiation crosses a layer in the plasma with critical density
\begin{equation}\label{eq:resdensity}
    \rho_\mathrm{V} \simeq 0.964 Y_\mathrm{e}^{-1}B_{14}^{2}E_{1}^{2}f(B)^{-2} \text{g cm}^{-3}\,;
\end{equation}
here $Y_\mathrm{e}$ is the electron fraction, $B_{14} = B/10^{14}
\mathrm{G}$, $E_1 = E/1\mathrm{keV}$ (with $E$ the photon energy) and $f(B)$ is a slow varying function of $B$ \citep{ho_ii_2003}. 
Near the resonant density, a photon which initially has an almost linear polarisation becomes circularly polarised and may then evolve to a linear polarisation state which is rotated by $90^\circ$. Namely, an X-mode photon can become an O-mode photon and vice versa, a phenomenon referred to as ``mode conversion'' \cite[see e.g.][]{ventura_scattering_1979}. 

In order to study the polarisation properties across the vacuum resonance, two limiting cases and one more general scenario are usually discussed in literature \citep{pavlov_effect_1979, ho_ii_2003, lai_transfer_2003, ho_iii_2003, gonzalez-caniulef_atmosphere_2019}. In the first limiting case, the photon ellipticities $K_+$ and $K_-$ evolve independently, simply following equation (\ref{elli}); in this case a photon initially polarised in the O- (X-)mode will convert to the X- (O-)mode (complete mode conversion). The opposite limiting case occurs when photons pass through the resonance remaining in their initial polarisation mode (no mode conversion). This means that the photon ellipticity switches between the $K_+$ and $K_-$ solutions of equation (\ref{elli}) at the resonance. In the general case (partial mode conversion), mode conversion occur only if some criteria are satisfied. These criteria depend  on both the photon energy and propagation direction with respect to the magnetic field.

Actually, it can be shown that, in the Wentzel–Kramers–Brillouin (WKB) approximation, the amplitudes $A_+$ and $A_-$ of the plus/minus modes evolve along the photon path according to
\begin{equation} \label{eqn:mixing}
    i\frac{d}{dz}\left(\begin{array}{c}
         A_+ \\ A_- 
    \end{array}\right)\simeq\left[\begin{array}{cc}
         -\Delta k/2 & i\theta_\mathrm{m} \\
         -i\theta_\mathrm{m} & \Delta k/2
    \end{array}\right]\left(\begin{array}{c}
         A_+ \\ A_-
    \end{array}\right)\,,
\end{equation}
where $\Delta k=k_+-k_-$ is the difference between the moduli of the wave vectors and $\theta_\mathrm{m}=\arctan{1/(2\beta)}$ is the mixing angle \citep{lai_resonant_2002}. From Equation (\ref{eqn:mixing}) it is clear that the photon ellipticity adiabatically follows the $K_+$ ($K_-$) curves, i.e. without mixing, if the off-diagonal terms in the matrix are much smaller than the diagonal ones. Evaluating the condition $\vert\theta_\mathrm m\vert\ll\vert \Delta k/2\vert$ at the resonance, $\rho=\rho_\mathrm V$, gives a limiting value of the energy above which propagation is adiabatic 
\begin{equation} \label{Eq:Ead}
    E_\mathrm{ad} = 2.52(f\tan\theta_\mathrm{Bk})^{2/3}\biggl(\frac{1\,\mathrm{cm}}{H_\rho}\biggl)^{1/3}\,\text{keV}\,,
\end{equation}
where $H_\rho$ is the density scale-height and $u_i\ll 1$ is assumed.
The probability that a photon ``jumps'' between the $+$ and $-$ modes (hence remaining in its original X or O polarisation mode), can be derived from the Landau-Zener formula  
\begin{equation} \label{Eq:Probability}
    P_\mathrm{J} = \exp{\biggl[-\frac{\pi}{2}\biggl(\frac{E}{E_\mathrm{ad}}\biggl)^3\biggl]}\,.
\end{equation}
In turn, the probability of mode conversion occurring is given by $P_\mathrm{con}=1-P_\mathrm{J}$. For $E=1.3E_\mathrm{ad}$, it is $P_\mathrm{J}\approx 0.03$, so that complete mode conversion is expected. 

\begin{figure}
    \centering
    \includegraphics[width=\columnwidth]{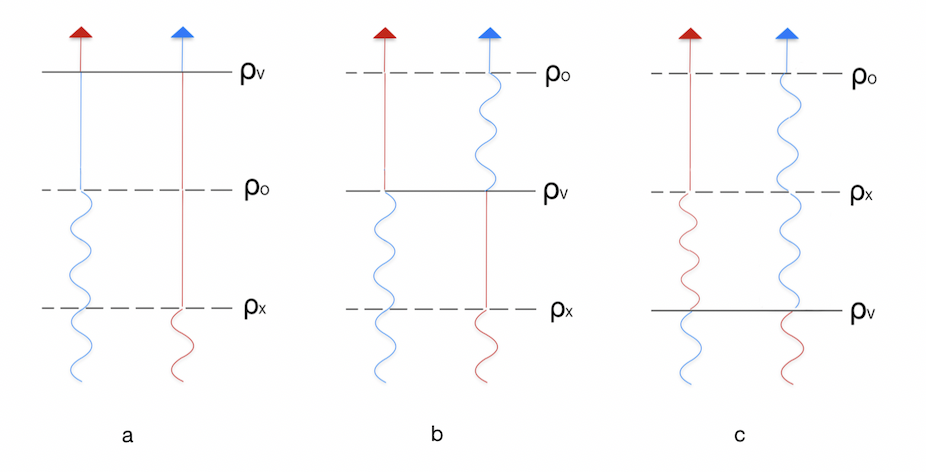}
    \caption{The effects on the emergent radiation of the different relative positions of the X/O mode photospheres (dashed lines) and the vacuum resonance (solid line); red (blue) are for X (O) mode photons, respectively. Straight (wiggly) lines show optically thin (thick) regions.
    }
    \label{fig:positions}
\end{figure}

The way in which mode conversion influences the polarisation of radiation propagating through the atmosphere can be understood retracing the argument presented by \citet{lai_ixpe_2023}. This basically depends on the relative positions of the X and O photospheres and of the vacuum resonance, as schematically illustrated in Figure \ref{fig:positions} assuming complete mode conversion. In the first scenario (Figure \ref{fig:positions}a), the vacuum resonance is at a density lower than that of the photosphere of both polarisation modes. Since X-mode photons decouple at higher densities (i.e. deeper in the atmosphere) there will be more X-mode than O-mode photons as radiation crosses the resonance. Here a large fraction of X-mode photons convert into O-mode photons and, with both modes decoupled, this results in radiation being O-mode dominated at the observer.

In the second case the resonance lies in between the two photospheres (see Figure \ref{fig:positions}b), with the X-mode photosphere deeper inside than the resonance. The flux before the resonance is dominated by X-mode photons, while O-mode ones prevail once the resonance is crossed. However, at this depth, the X-mode photons are free to travel to the observer while the O-mode ones are still trapped and continue to interact, potentially changing into the X-mode, until they eventually reach the O-mode photosphere, resulting in a spectrum that therefore remains dominated by X-mode photons.

Finally, if the vacuum resonance is at a higher density (deeper in the atmosphere) than that of both polarisation mode photospheres (Figure \ref{fig:positions}c), mode conversion has no effect since both X- and O-mode photons are in thermal equilibrium either before and after crossing the resonance. X-mode photons then decouple deeper in the atmosphere, at a higher depth, than the O-mode photons and the emission remains X-mode dominated.

In the following we present atmospheric calculations and compute the expected spectra and polarisation by investigating the case of complete, no, and partial mode conversion. In the latter case we treat the probability using a reformulation of Eq. (\ref{Eq:Probability}) suitable for inclusion in our radiative transfer finite difference code.

\section{Numerical setup}
\label{sec:code} 

For producing our atmosphere models, we re-adapted the code by \citet{lloyd_model_2003}, designed for computing the radiation transport in a plane-parallel, geometrically thin atmospheric layer with a fully-ionised hydrogen composition. Each atmosphere model is characterised by  the magnetic field strength and inclination with respect to the surface normal, the surface gravity and the effective temperature.

Once a trial profile of the temperature $T$ and density $\rho$ as a function of depth is established, the code computes the monochromatic opacity
\begin{equation} \label{Eq:chi}
    \chi_\nu^j = \frac{\alpha_j^\mathrm{ff}+\alpha_j^\mathrm{sc}}{\rho}
\end{equation}
and  emissivity
\begin{equation} \label{Eq:eta}
    \eta_\nu^j = \frac{\epsilon_j^\mathrm{ff}+\epsilon_j^\mathrm{sc}}{\rho}\,,
\end{equation}
including both thermal free-free (ff) and Thomson scattering (sc); here $j$ stands for each of the two normal polarisation modes and $\alpha$ ($\epsilon$) is the monochromatic absorption (emission) coefficient. It then 
solves the radiative transfer equations, 
\begin{equation} \label{Eq:RadTran}
    \mu\frac{dI_\nu^j(\boldsymbol k)}{\rho dz} = \chi_\nu^j(\boldsymbol k)I_\nu^j(\boldsymbol k) - \eta_\nu^j( \boldsymbol k)\,,
\end{equation}
upon writing them in terms of Feautrier variables, on a energy, angular and depth mesh; the number of grid points and their spacing can be adjusted to improve convergence. In equations (\ref{Eq:RadTran}), $I_\nu^j$ is the monochromatic intensity for mode $j$, $z$ is the vertical coordinate  and $\boldsymbol k$ is the photon momentum, specified by the angles $\theta_\mathrm k$, that the ray makes with the $z$-axis, and $\phi_\mathrm k$, the associated azimuth. Actually, the latter angle is required only when $\boldsymbol B$ is not aligned with the local normal (i.e. $\theta_\mathrm B\neq 0$) since in these cases radiative transfer is inherently 3D in the photon momentum space because of the preferential direction introduced by the magnetic field. 
Standard boundary conditions were applied: no external illumination at the top of the atmosphere and a thermalisation condition for each mode at the bottom, where the radiation field is nearly isotropic, and  $I_\nu^j\simeq B_\nu/2$.

The radiative transfer equation  is solved numerically alongside hydrostatic and radiative equilibrium conditions and the energy balance equation on a depth, angle and frequency mesh. A complete-linearisation technique is applied and the temperature profile correction is 
computed  using the Uns\"{o}ld-Lucy method \cite[see e.g.][]{lucy_temperature-correction_1964, unsold_physik_1955, mihalas_stellar_1978} to ensure flux conservation at all depths. The entire procedure is repeated through successive iterations, 
until fractional corrections on temperature and flux drop below a prescribed accuracy (typically $10^{-3}$ in our runs).

The code can use a modal description based either on the $(\boldsymbol e_\mathrm X\,, \boldsymbol e_\mathrm O)$ or the $(\boldsymbol e_\mathrm +\,, \boldsymbol e_\mathrm -)$ basis. The appropriate choice is dictated by the assumed prescription for mode conversion. If no mode conversion occurs radiative transfer is solved directly in terms of the X/O modes. On the other hand, models with complete (or partial, see below) 
mode conversion are computed by solving the 
radiative transfer for the $\pm$ modes. In fact, for complete mode conversion, the $\pm$ modes remain unchanged across the vacuum resonance (and at the same time X-mode photons turn into O-mode and viceversa).
To simulate partial, adiabatic mode conversion, we prescribe in input a (fixed) value for the conversion probability $\bar P_\mathrm{con}$ and derive the corresponding energy threshold by inverting equation (\ref{Eq:Probability})
\begin{equation} \label{Econ}
    \bar E_\mathrm{con} = \biggl[-\frac{2}{\pi}\ln(\bar P_\mathrm{con})\biggl]^{1/3}E_\mathrm{ad}\,.
\end{equation}
If the photon energy in a given bin exceeds $\bar E_\mathrm{con}$ then the $\pm$ modes remain unchanged at the resonance and mode conversion occurs for the X/O ones. In the opposite case, the $\pm$ modes switch, i.e. $K_\pm\to K_\mp$ at the resonance. After the resonance, we recover the X/O modal description by associating the O-mode to the rays for which 
$|K_-|\gg1$ and the X-mode to those with $|K_+|\ll1$.
While doing  this, we evaluate the quantity 
$E_\mathrm{ad}$ at a  reference value of $\theta_\mathrm{Bk}$, which we take as $\tan \theta_\mathrm{Bk} =1$ (but see the end of \S~\ref{sec:results} for a discussion on the impact of this assumption). 
We notice that this is a simplified approach since the conversion probability should be computed, at a given energy, ray-by-ray, because $E_\mathrm{ad}$ depends on $\theta_\mathrm{Bk}$ and hence on $\theta_\mathrm k$, $\phi_\mathrm k$ and $\theta_\mathrm B$. Implementing the proper algorithm, even if possible at all, would require a major upgrade
of our code because of the way the scattering integrals are calculated.

\section{Results} \label{sec:results}

In the following we explore how vacuum corrections, assuming either complete or partial mode conversion, impact on the polarisation properties of the radiation emerging from a magnetar atmosphere, allowing for different effective temperatures and magnetic field strengths and inclinations. We also investigate the effect of different conversion criteria on the partial mode conversion models. We define the polarisation degree of the emergent radiation as
\begin{equation} \label{PD}
    \mathrm{PD}_\mathrm{em} = \frac{J_\nu^\mathrm{X} - J_\nu^\mathrm{O}}{J_\nu^\mathrm{X} + J_\nu^\mathrm{O} }\,,
\end{equation}
so that positive polarisation degrees indicate X-mode dominated spectra; here $J_\nu^{j}$ denotes the mean intensity of mode $j$.

\begin{figure*}
  \centering
  
  \subfigure[]{\includegraphics[width=\columnwidth]{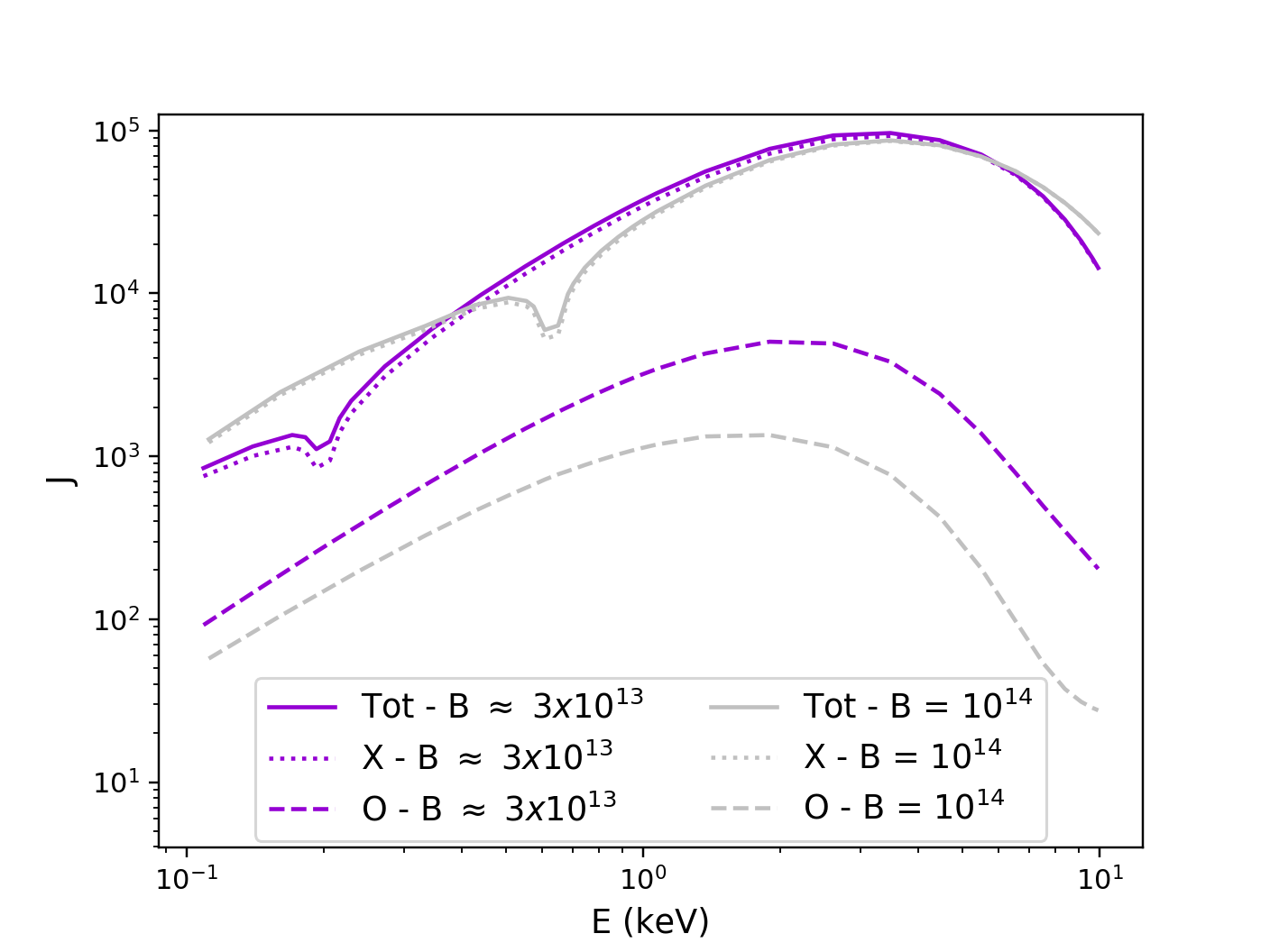}}
  \subfigure[]{\includegraphics[width=\columnwidth]{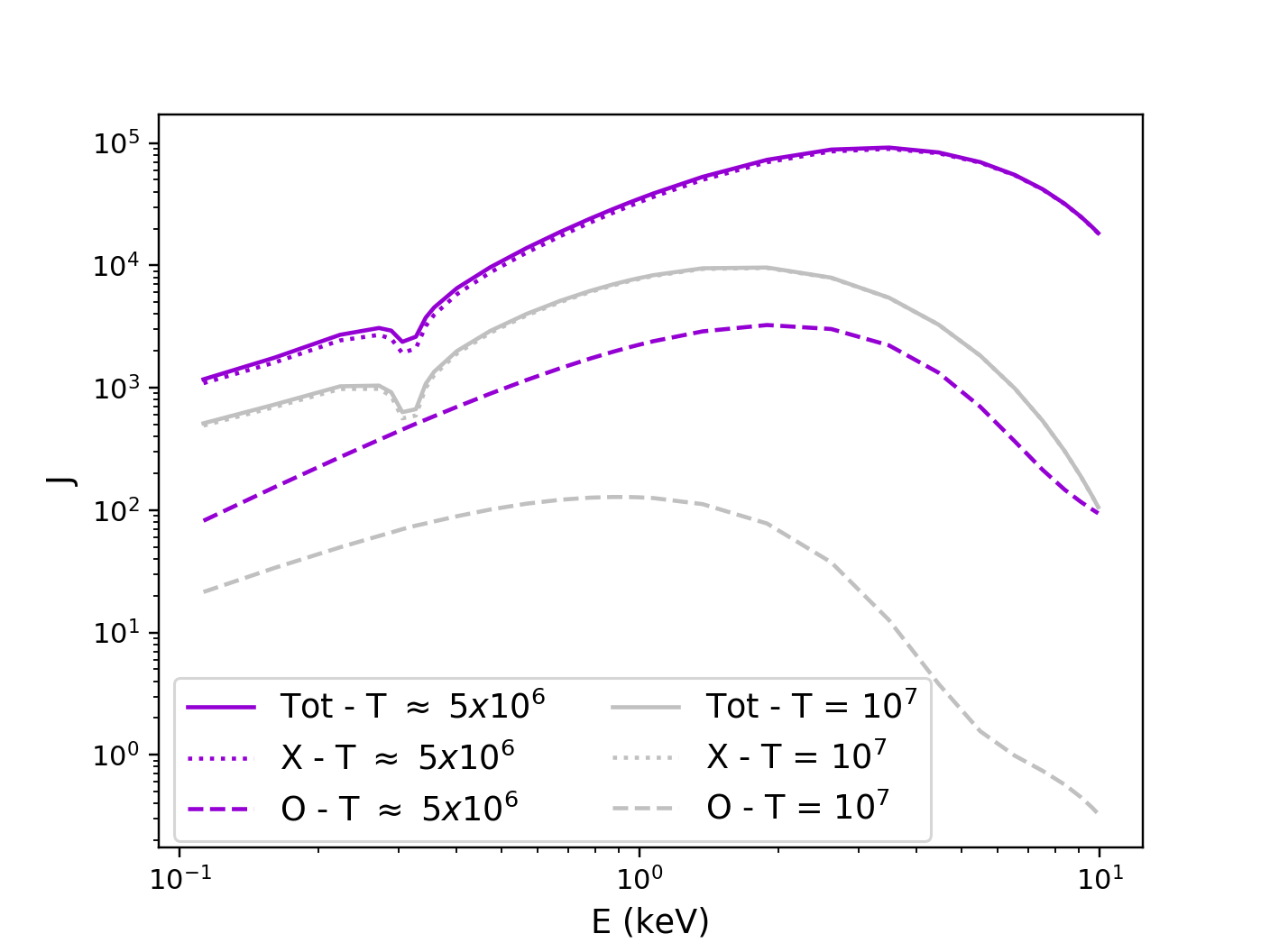}}  
  \subfigure[]{\includegraphics[width=\columnwidth]{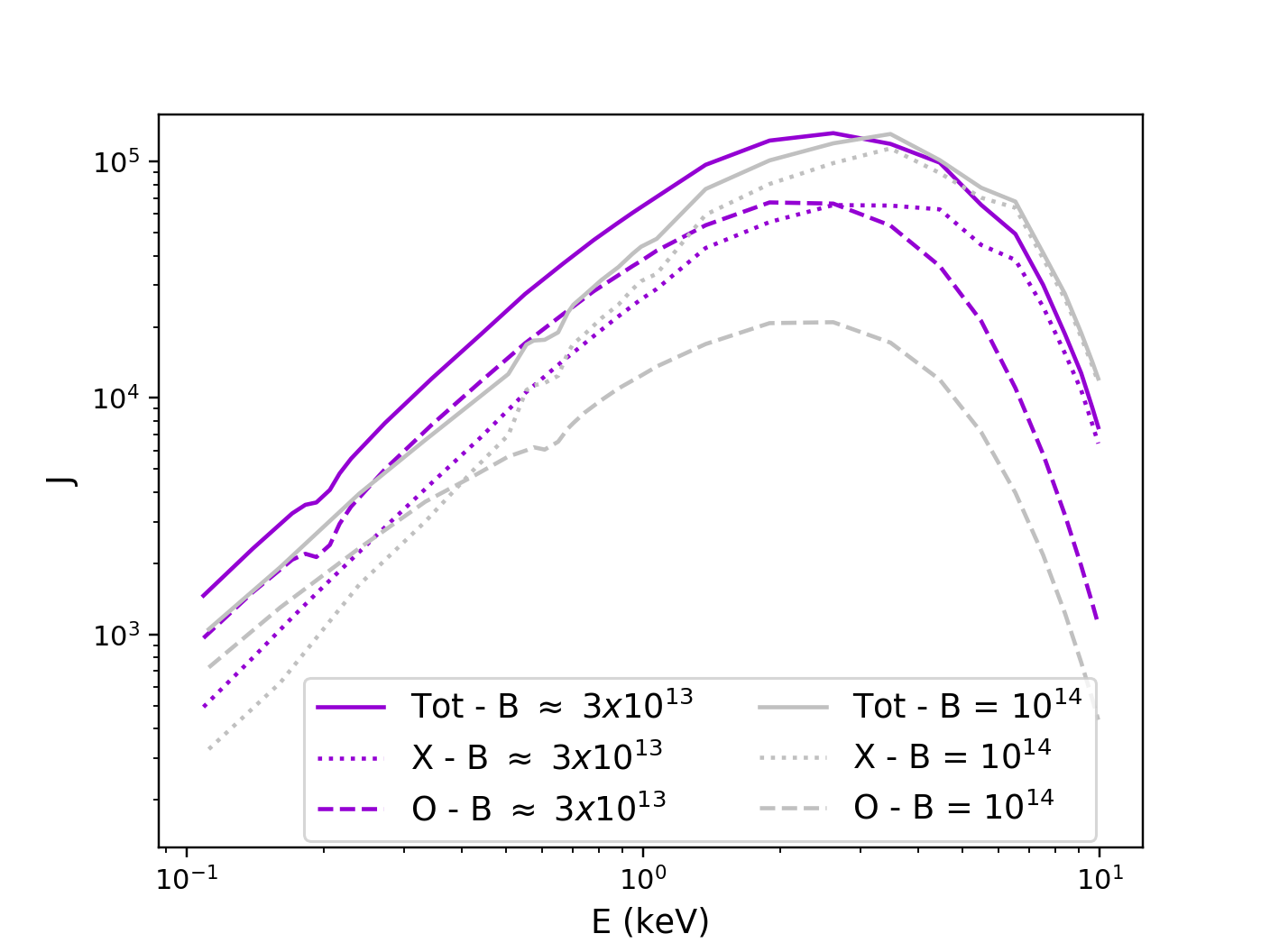}}
  \subfigure[]{\includegraphics[width=\columnwidth]{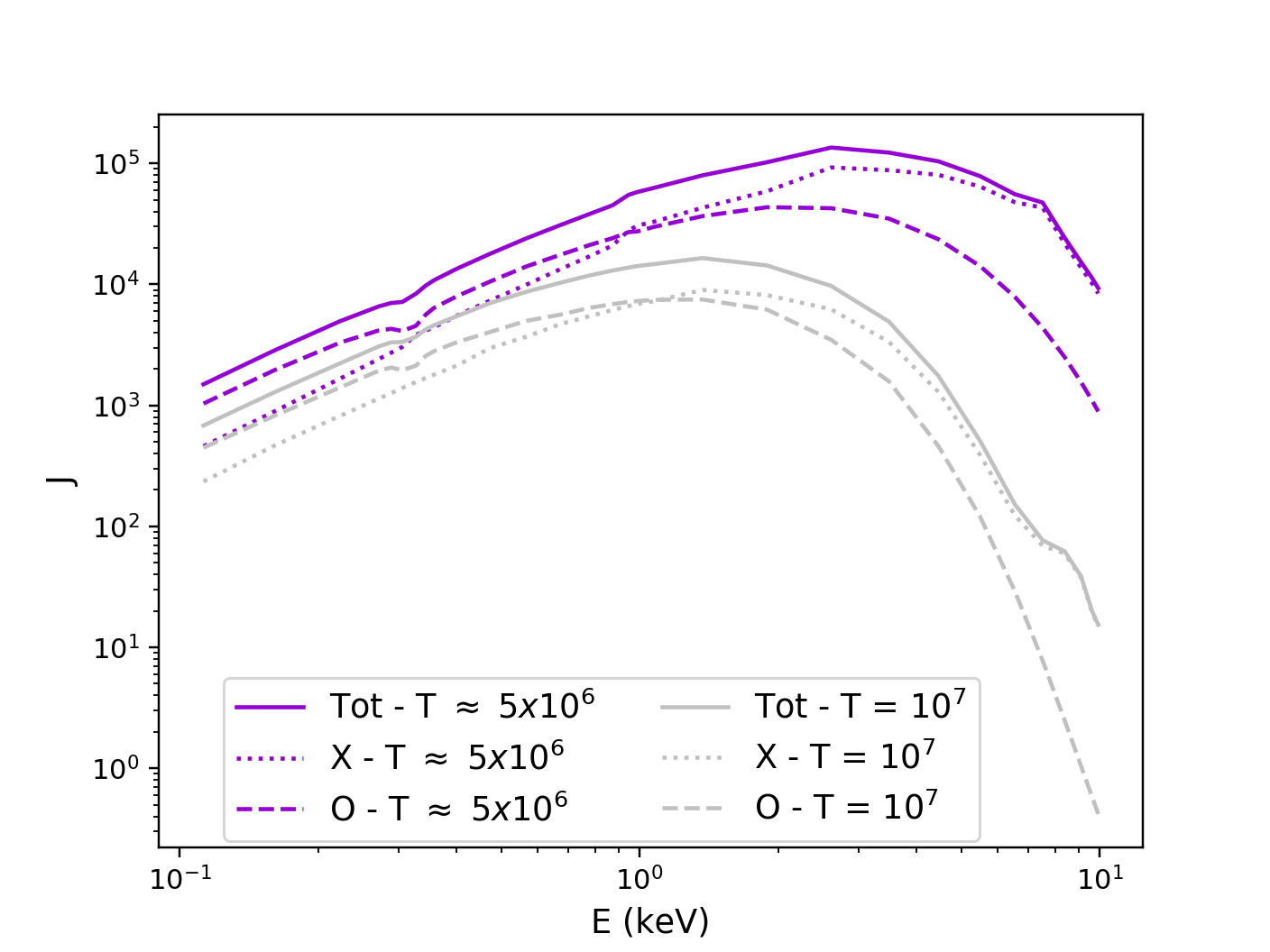}}
  \subfigure[]{\includegraphics[width=\columnwidth]{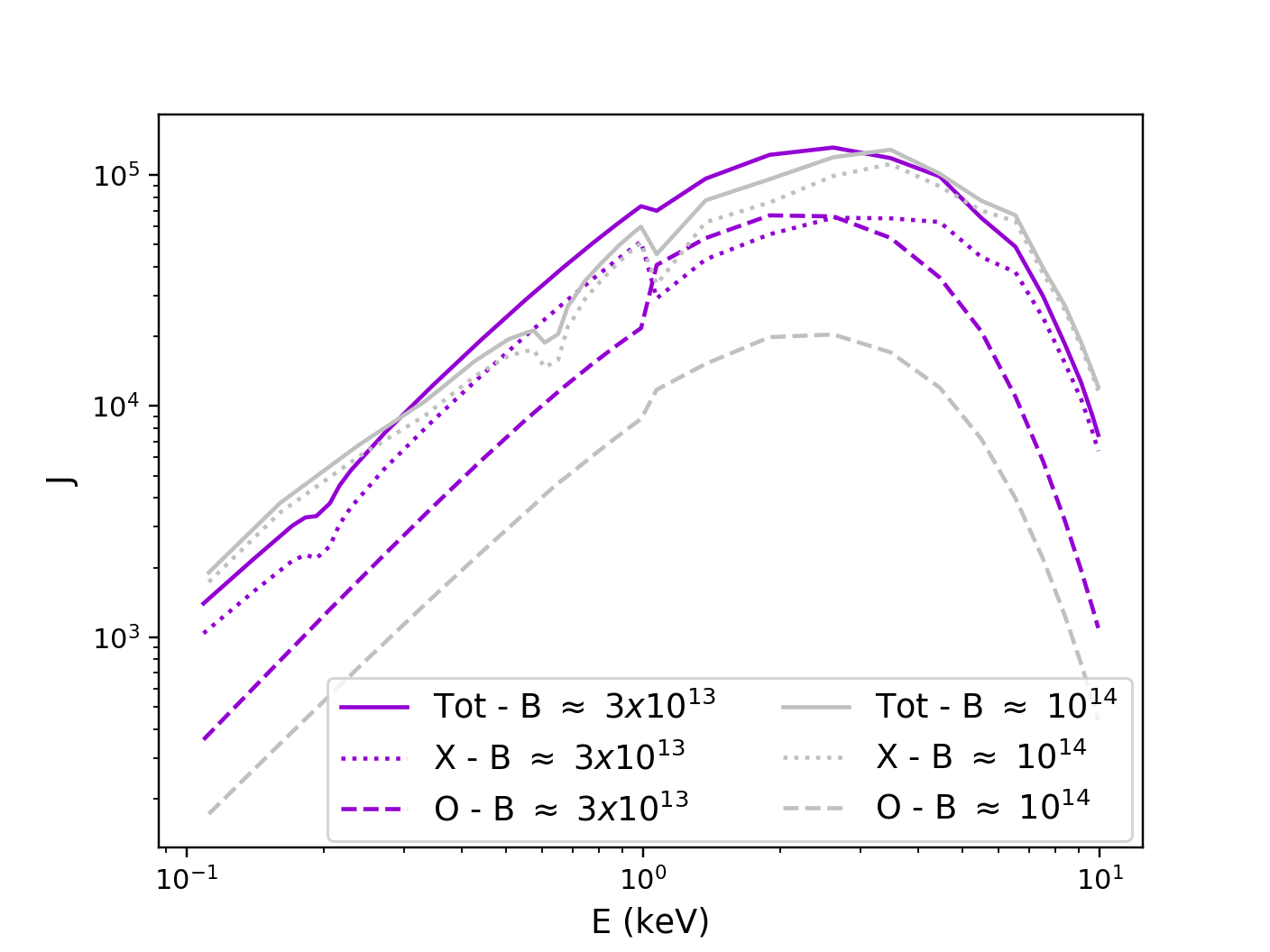}}
  \subfigure[]{\includegraphics[width=\columnwidth]{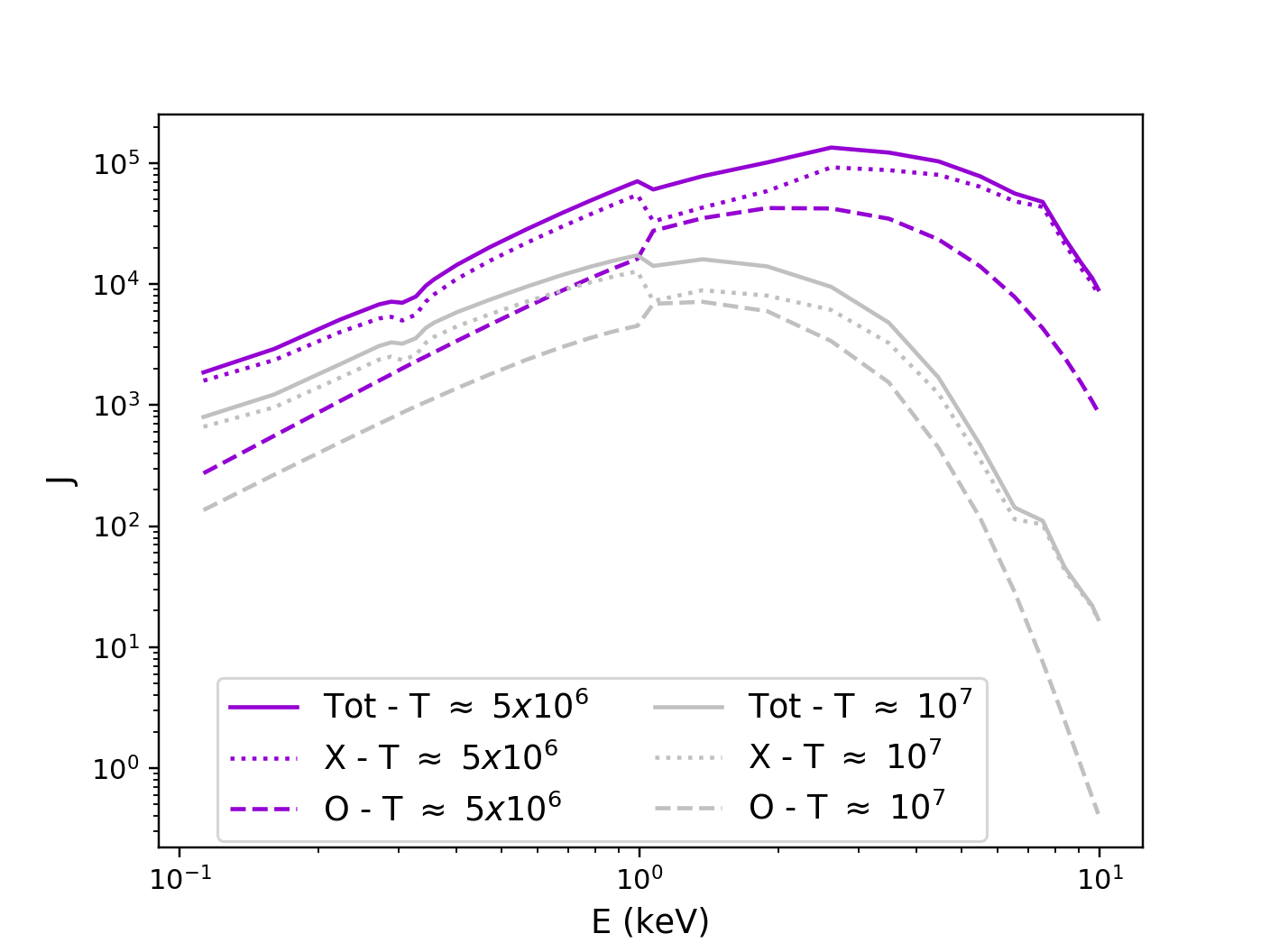}}
  
  \caption{Mean intensity spectrum of the emergent radiation in the soft X-ray band as a function of the photon energy $E$ in the cases of a standard pure plasma atmosphere without vacuum corrections (panels a and b), complete mode conversion (panels c and d) and partial, adiabatic mode conversion with $P_\mathrm{con}=0.1$ (panels e and f). In the left-hand column models with different magnetic field strengths at the same effective temperature $T=10^7\,\mathrm{K}$ and magnetic field inclination $\theta_\mathrm B=0^\circ$ are reported, while the right-hand one shows models with different effective temperatures at fixed magnetic field strength  $B\approx5\times10^{13}\,\mathrm{G}$ and inclination $\theta_\mathrm B=0^\circ$. Here, solid, dotted and dashed lines show the total, X-mode and O-mode mean intensities, respectively.}
  \label{fig:intensity}
\end{figure*}

Figure \ref{fig:intensity} shows the emergent mean intensity from atmospheric models computed assuming no vacuum corrections or with vacuum corrections and complete or partial mode conversion at the vacuum resonance. The models in the left-hand column refer to an effective temperature of $T = 10^{7}$~K and different magnetic field strengths, while those in the right-hand one refer to a fixed magnetic field strength of $B = 5\times10^{13}$~G and different effective temperatures. In all cases the magnetic field is assumed to be aligned to the surface normal $\boldsymbol n$ (i.e. $\theta_\mathrm{B}=0^\circ$). For models computed assuming a standard, pure-plasma atmosphere, without vacuum corrections, the emergent X-mode intensity is almost coincident with the total intensity while the O-mode intensity is significantly lower. As it is well known, this is due to the higher opacity experienced by O-mode photons as they travel through the atmospheric plasma. Additionally, a very clear absorption line can be seen in the spectra (Figure \ref{fig:intensity}a) in correspondence of the proton cyclotron energy $E_\mathrm{Bp} \simeq 0.63 (B/10^{14}\,\mathrm{G})\,\mathrm{keV}$ \citep{zane_proton_2001}.

Models computed with the vacuum contributions and assuming complete mode conversion (Figure \ref{fig:intensity}c and d) paint a very different picture with respect to the previous case. In these models, obtained for the same parameters, the X- and O-mode intensities are much closer together, with the O-mode having a higher intensity than the X-mode in the low energy range while the opposite occurs at  higher energies.
This is expected as low energy photons encounter the vacuum resonance higher up in the atmosphere (i.e at lower density) with respect to the O-mode photosphere. Therefore, at the vacuum resonance, photons with both polarisation modes are decoupled from the plasma, and what was a large amount of outgoing X-mode photons convert into O-mode photons (and vice versa a small amount of O-mode photons convert to X-mode, see Figure \ref{fig:positions}a), resulting in an emergent spectrum dominated by O-mode photons. On the other hand,  photons with higher energy encounter the vacuum resonance deeper in the atmosphere, eventually inside the O-mode photosphere, therefore the emergent spectrum appears X-mode dominated at high energy (see Figure \ref{fig:positions}b and c). It is worth noticing that 
the energy at which the dominant polarisation mode shifts from O to X decreases with increasing magnetic field strength (as can be seen from Figure \ref{fig:intensity}c), due to the increase in the density at which the vacuum resonance occurs (Equation \ref{eq:resdensity}). For instance, for $B= 3 \times 10^{13}$~G the shift occurs at $E\approx 2.5$~keV, while for $B= 10^{14}$~G it is well below $2\, \mathrm{keV}$, at $E\approx 0.4\, \mathrm{keV}$. 
Interestingly, as also discussed by \citet{zane_proton_2001}, \citet{ho_iii_2003} and \citet{van_adelsberg_atmosphere_2006}, the depth of the proton cyclotron line is significantly reduced  due to the contribution of vacuum polarisation. The decrease becomes more pronounced as the magnetic field strength increases. We also notice that there is a slight decrease in the O-mode intensity at the same energy, possibly resulting from some cyclotron absorption of X-mode photons which, in turn, reduces the intensity of the photons which are then converted to O-mode photons at the vacuum resonance.

In Figure \ref{fig:intensity}e and f the mean intensity for models computed including partial adiabatic mode conversion are shown, using the same atmosphere parameters as above to ease comparison. Now at low energies the X-mode intensity is significantly higher than the O-mode one, and the difference between the two intensities becomes larger with higher magnetic field strengths. The proton cyclotron line is also clearly present, although it again appears reduced in depth as a result of the vacuum contribution.
In these models, the probability criterion for mode conversion (chosen to be $P_\mathrm{con}=0.1$ in this case) is met at an energy $E_\mathrm{con}\approx1$~keV: above this energy mode conversion at the vacuum resonance occurs for every photon, and this translates into a sudden decrease in the intensity of the X-mode and increase in the O-mode intensity.

Interestingly, as can be seen in Figure \ref{fig:intensity}e, for moderate field strengths, $B\approx3\times10^{13}$ G, there are now two values of the energy at which the dominant emergent polarisation mode changes: the first switch occurs at $E_\mathrm{con}$, above which the O-mode intensity becomes higher than the X-mode, and the second around $2$--$3$ keV. The latter corresponds to the switch we previously discussed for the case of complete mode conversion (Figure \ref{fig:intensity}c) of the same magnetic field strength, and is the result of the vacuum resonance moving within the O-mode photosphere in the atmosphere.
On the other hand, for higher magnetic field strengths, $B\approx 10^{14}$~G, the X-mode photon intensity remains dominant throughout the entire energy range. This is again due to the relative positions of the mode photospheres and the vacuum resonance: for magnetic fields $\gtrsim5\times10^{13}\,\mathrm{G}$, photons with energy $\gtrsim E_\mathrm{con}$ encounter the vacuum resonance in deeper atmospheric layers, within the O-mode photosphere, and the spectrum therefore remains X-mode dominated throughout. Conversely, for fields  $\lesssim 5\times10^{13}\,\mathrm{G}$, photons with energy $\approx E_\mathrm{con}$ encounter the vacuum resonance higher in the atmosphere and outside the O-mode photosphere: this translates in a spectrum that can be dominated by the O-mode for some part of the energy range, resulting in two values of the energy  at which the higher intensity mode switches. Additionally, spectra exhibit a reduction in the total intensity at energies near $E_\mathrm{con}$. This feature is however not a physical absorption feature: we tested that it is the result of the assumption we used for the mode-conversion probability (approximated as a step function), and would likely not be present if the probability was instead treated as a smooth function.
For all three groups of models, a decrease in the effective temperature (keeping all other parameters fixed) from $10^7$~K to $5 \times 10^6$~K  simply results in a decrease in the total intensity, as well as in single the X- and O-mode components, but the features discussed so far remain present (Figure \ref{fig:intensity}, right-hand column).

\begin{figure}
    \centering
    \includegraphics[width=\columnwidth]{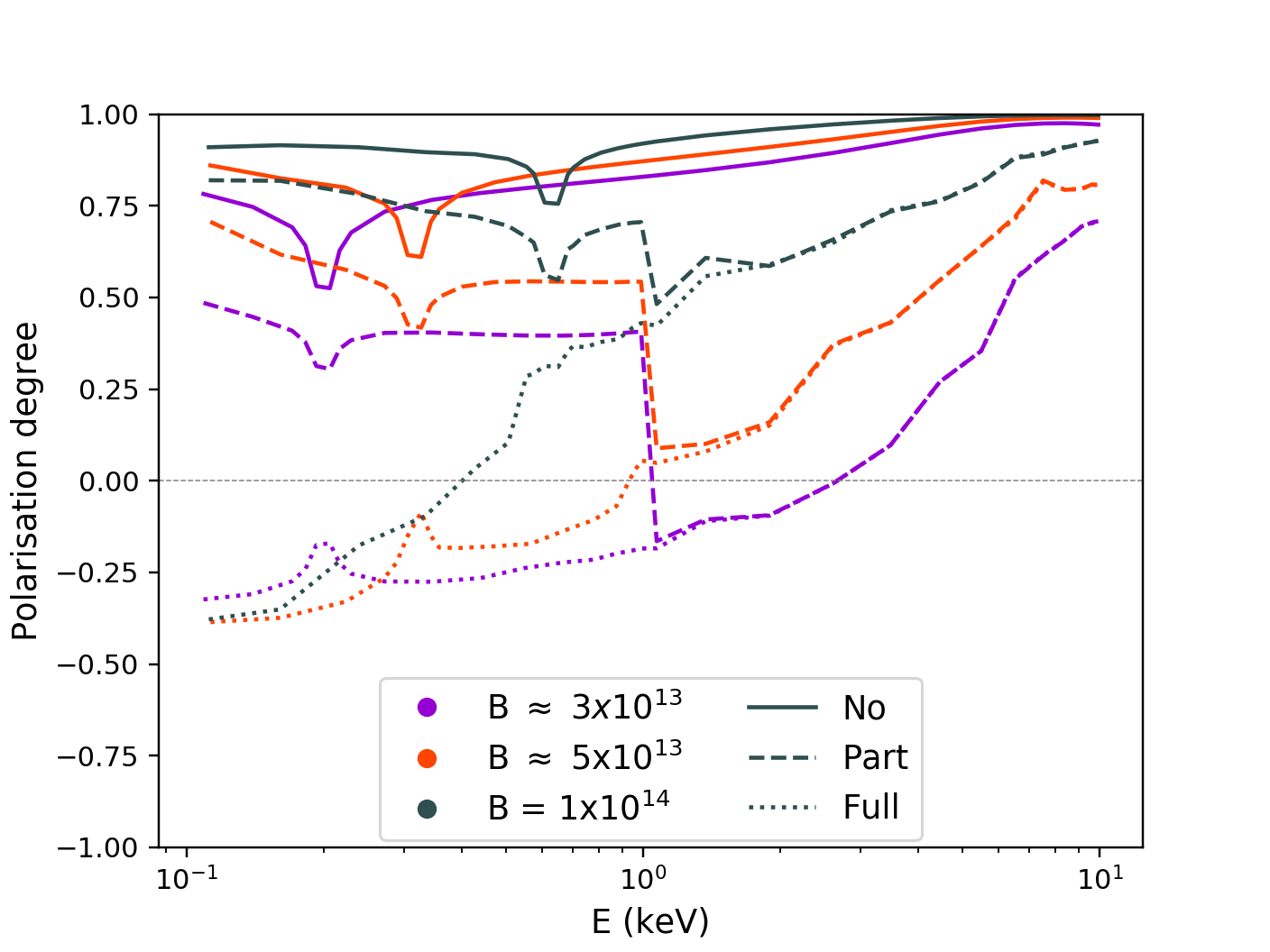}
    \caption{Polarisation degree of the emergent radiation as a function of the photon energy $E$ for different magnetic field strengths, $B\approx3\times10^{13}\,\mathrm{G}$, $5\times10^{13}\,\mathrm{G}$ and $10^{14}\,\mathrm{G}$, with fixed effective temperature $T=10^7$ K and magnetic field inclination $\theta_\mathrm{B}=0^\circ$. Here the case of an atmosphere with no vacuum corrections is marked by solid lines, dashed lines denote models of atmospheres with partial adiabatic mode conversion 
    and dotted lines refer to complete mode conversion. See text for more details.}
    \label{fig:Bfields}
\end{figure}

Figure \ref{fig:Bfields} shows the effect of varying the magnetic field strength on the polarisation degree of the emergent radiation for a magnetar with an effective temperature $T = 10^{7}$ K; we also assumed a magnetic field inclination $\theta_\mathrm{B}=0^\circ$, 
and $P_\mathrm{con}=0.1$.
As expected, without including vacuum corrections the emergent signal shows an increase in polarisation with increasing magnetic field strength. Above the proton cyclotron line, the polarisation also increases with the photon energy and tends towards $100\%$ polarisation in the X-mode at $E \sim 10$ keV. Conversely, 
computations carried out assuming complete mode conversion result in O-mode dominated emergent radiation at low energies ($\sim 0.1$ keV), due to the relative positions of the vacuum resonance and the photospheres, with a typical polarisation degree of $\approx 20$--$30\%$. The energy at which the dominant polarisation mode switches from O to X decreases with increasing magnetic field strength, as a result of the increasing resonance density (see Equation \ref{eq:resdensity}). Above this energy, the polarisation degree generally increases (in the X-mode) up to $\approx 60$--$80\%$ at $\sim 10\,\mathrm{keV}$, attaining higher values for larger magnetic field strengths.

Finally, Figure \ref{fig:Bfields} also shows 
the polarisation properties for the partial mode conversion models. In this case, photons with $E < E_\mathrm{con}$ (which turns out to be $\sim 1$~keV for the parameters used here) have not met the conversion probability criterion and therefore do not convert from one mode to the other at the vacuum resonance. The polarisation degree of the emergent radiation, at these energies,
increases with the magnetic field strength, similar to what is seen in the no-mode conversion model. However, the polarisation degree in the case of partial mode conversion is reduced by $\sim 20\%$  when compared with the pure-plasma atmosphere models, due to the contribution of vacuum corrections.
At $E \approx E_\mathrm{con}$, the probability criterion is met and photons do convert from one polarisation mode to the other as they pass through the vacuum resonance, resulting in a significant change in the polarisation degree of the emergent radiation.
The severity of the sudden decrease in polarisation degree occurring at $E_\mathrm{con}$ depends on the magnetic field strength (with higher magnetic field strengths resulting in smaller decreases), again due to the relative positions of the vacuum resonance with respect to the photospheres. Finally, at energies above $E_\mathrm{con}$, the polarisation computed in the partial mode conversion cases is in almost complete agreement with the polarisation from the complete mode conversion ones with the same field strength, as expected.

\begin{figure}
    \centering
    \includegraphics[width=\columnwidth]{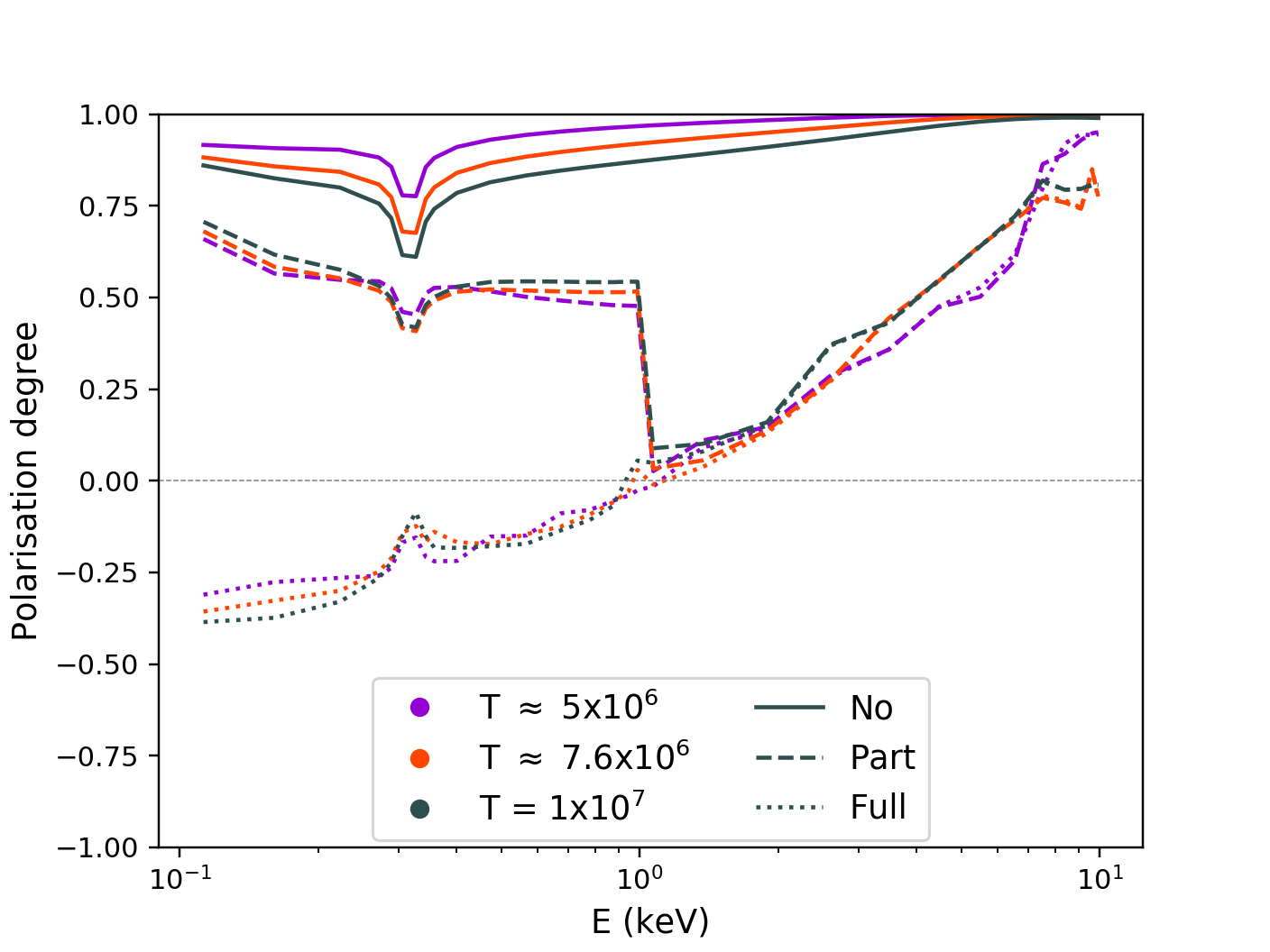}
    \caption{Polarisation degree of the emergent radiation as a function of the photon energy $E$ for different effective temperatures, $T\approx5\times10^{6}\,\mathrm{K}$, $7.6\times10^{6}\,\mathrm{K}$ and $10^{7}\,\mathrm{K}$, with fixed magnetic field strength $B\approx5\times10^{13}\,\mathrm{G}$ and inclination $\theta_\mathrm{B}=0^\circ$; the line styles are the same as in Figure \ref{fig:Bfields}.}
    \label{fig:Temp}
\end{figure}

The effect on the polarisation degree of the effective temperature is shown in Figure \ref{fig:Temp}. The models were produced with $B\approx5\times10^{13}\, \mathrm G$ and $\theta_\mathrm{B}=0^\circ$. The polarisation degree of a standard, pure plasma atmosphere decreases slightly as the effective temperature increases.
Models computed assuming complete mode conversion have a similar trend with temperature in the lower energy range ($\sim0.11$--$1\,\mathrm{keV}$). However, as the energy increases, this trend becomes less evident.
Interestingly, this effect is less pronounced in the case of partial-mode-conversion atmospheres, which shows the polarisation degree marginally increasing with increasing effective temperature.

\begin{figure}
    \centering
    \includegraphics[width=\columnwidth]{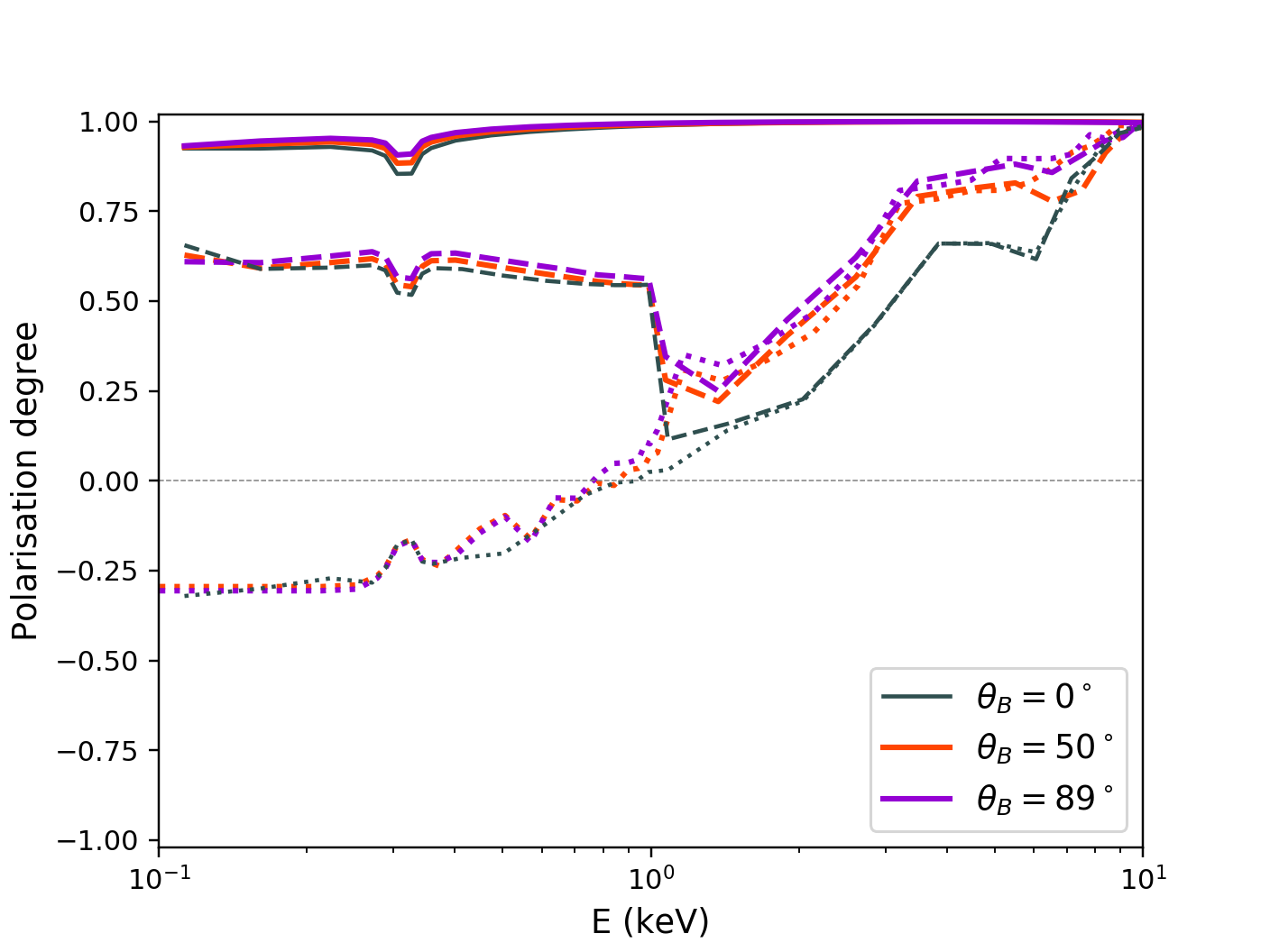}
    \caption{Polarisation degree of the emergent radiation as a function of the photon energy $E$ for different magnetic field inclinations $\theta_\mathrm{B}=0^\circ$, $50^\circ$ and $89^\circ$, with fixed effective temperature $T=5\times10^6\, \mathrm K$  and magnetic field strength $B\approx5\times10^{13}\, \mathrm G$. Line styles represent the different mode-conversion scenarios as in Figures \ref{fig:Bfields} and \ref{fig:Temp}.}
    \label{fig:inclined}
\end{figure}

From Figure \ref{fig:inclined}, the effect of the magnetic field inclination on the emergent polarisation from an atmosphere with $T = 5\times10^{6}\,\mathrm K$, $B\approx5\times10^{13}\, \mathrm G$, and  $P_\mathrm{con} = 0.1$ 
can be seen. As expected, for complete mode conversion, the inclination angle has little effect on the energy at which the switch in dominant polarisation mode (O to X) occurs. However, for both complete and partial mode conversion, at higher energies (above $\sim 1 \, \mathrm{keV}$) the polarisation degree is increased slightly for inclined cases ($\theta_\mathrm{B}=50^\circ$ and $\theta_\mathrm{B}=89^\circ$) with respect to the $\theta_\mathrm{B}=0^\circ$ case.

\begin{figure}
    \centering
    \includegraphics[width=\columnwidth]{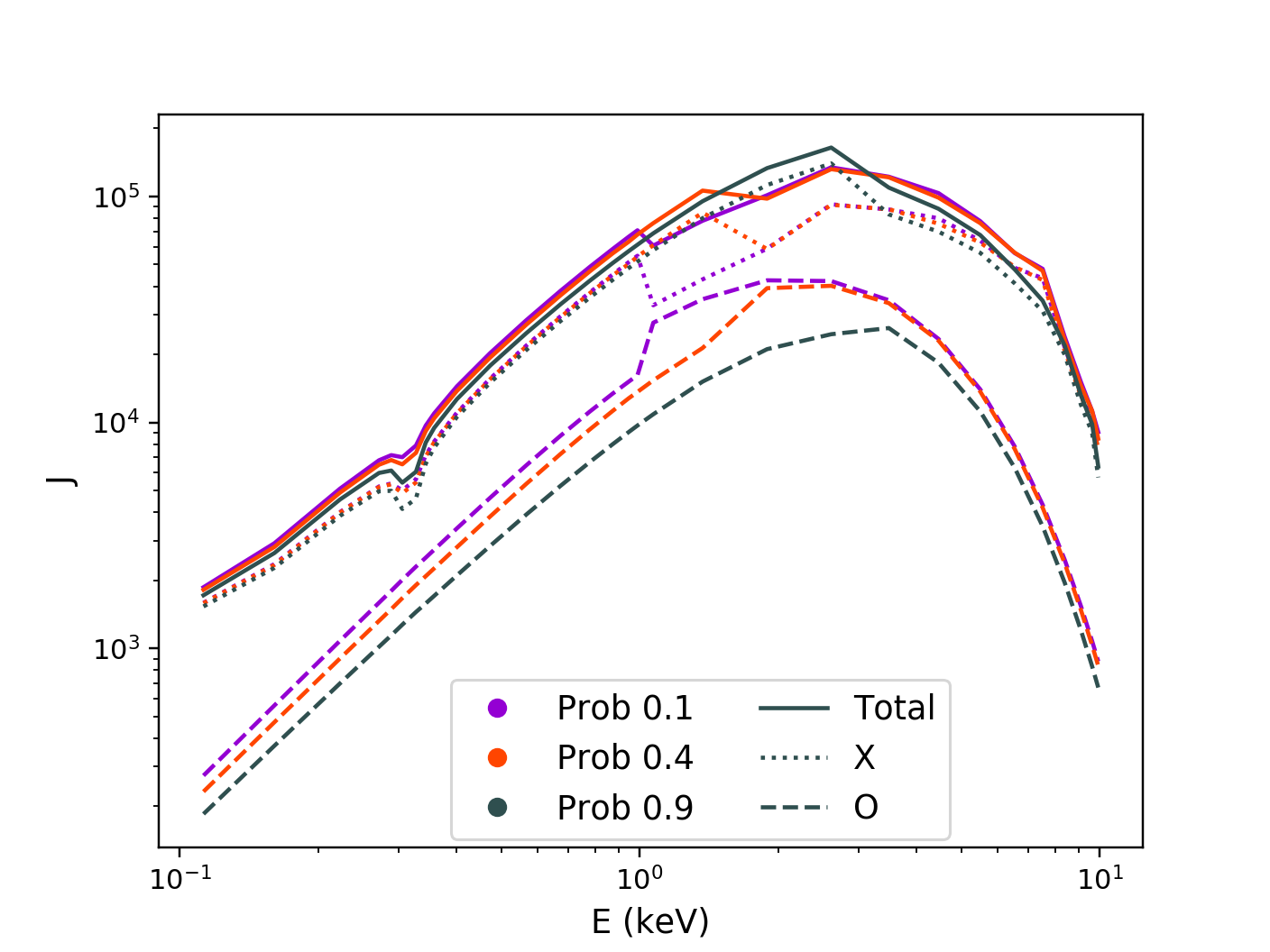}
    \caption{Mean intensity spectrum of the emergent radiation, for partial mode conversion models with different assumptions in the probability thresholds. Here $T=10^7\, \mathrm K$, $B\approx5\times10^{13}\, \mathrm G$ and $\theta_\mathrm{B}=0^\circ$. Solid, dotted and dashed lines show the total, X-mode and O-mode intensity respectively.}
    \label{fig:intenprob}
\end{figure}

Finally, for completeness, we investigated the effect of our assumption on the probability criteria for mode conversion by computing models with  different values for the threshold $P_\mathrm{con}$, at which we assume that  mode conversion occurs. 
Results are shown in Figure~\ref{fig:intenprob} for models with $T=10^7\, \mathrm K$, $B\approx5\times10^{13}\, \mathrm G$ and $\theta_\mathrm{B}=0^\circ$.
As expected, the only significant change is in the value of the energy at which the X-mode intensity decreases and the O-mode intensity increases, i.e. $E_\mathrm{con}$.
When the assigned threshold probability is higher, so too is the photon energy at which mode conversion occurs. Hence, for magnetic field strengths $\gtrsim 5 \times 10^{13}$ G, increasing the threshold value results in a larger difference between the X-mode and O-mode intensities across the entire energy range.

\begin{figure}
    \centering
    \includegraphics[width=\columnwidth]{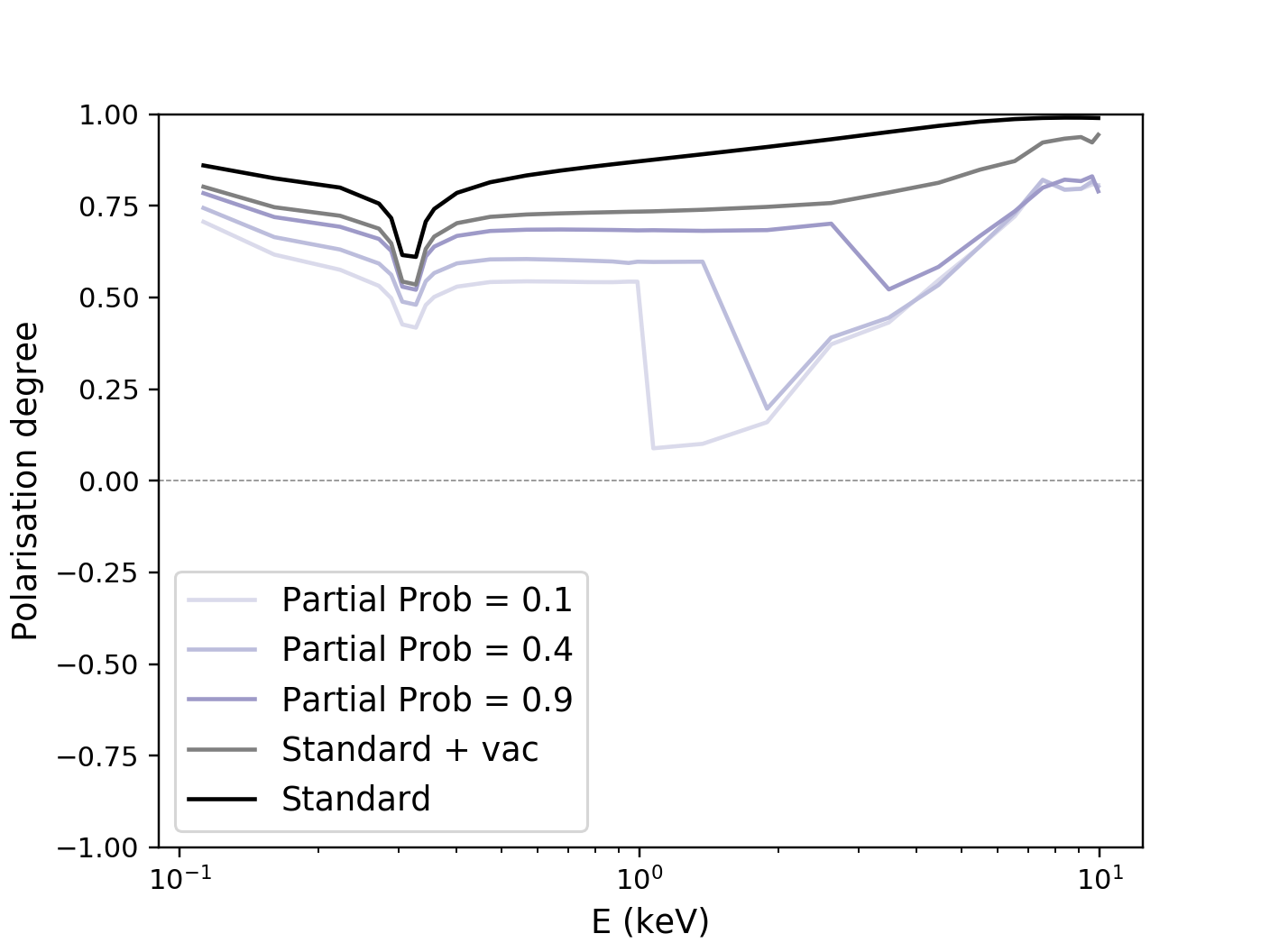}
    \caption{Polarisation degree of the emergent radiation computed assuming  partial adiabatic mode conversion, as a function of the photon energy $E$, for different assumptions on the probability thresholds, with fixed effective temperature $T=10^7\mathrm K$, magnetic field strength $B \approx 5\times10^{13}\, \mathrm G$ and inclination $\theta_\mathrm{B}=0^\circ$. The polarisation degree of the emergent radiation for an atmosphere computed without (termed standard in the legend) and with vacuum corrections but assuming no mode conversion (standard+vac in the legend) is also shown.}
    \label{fig:Prob}
\end{figure}

\begin{figure}
    \centering
    \includegraphics[width=\columnwidth]{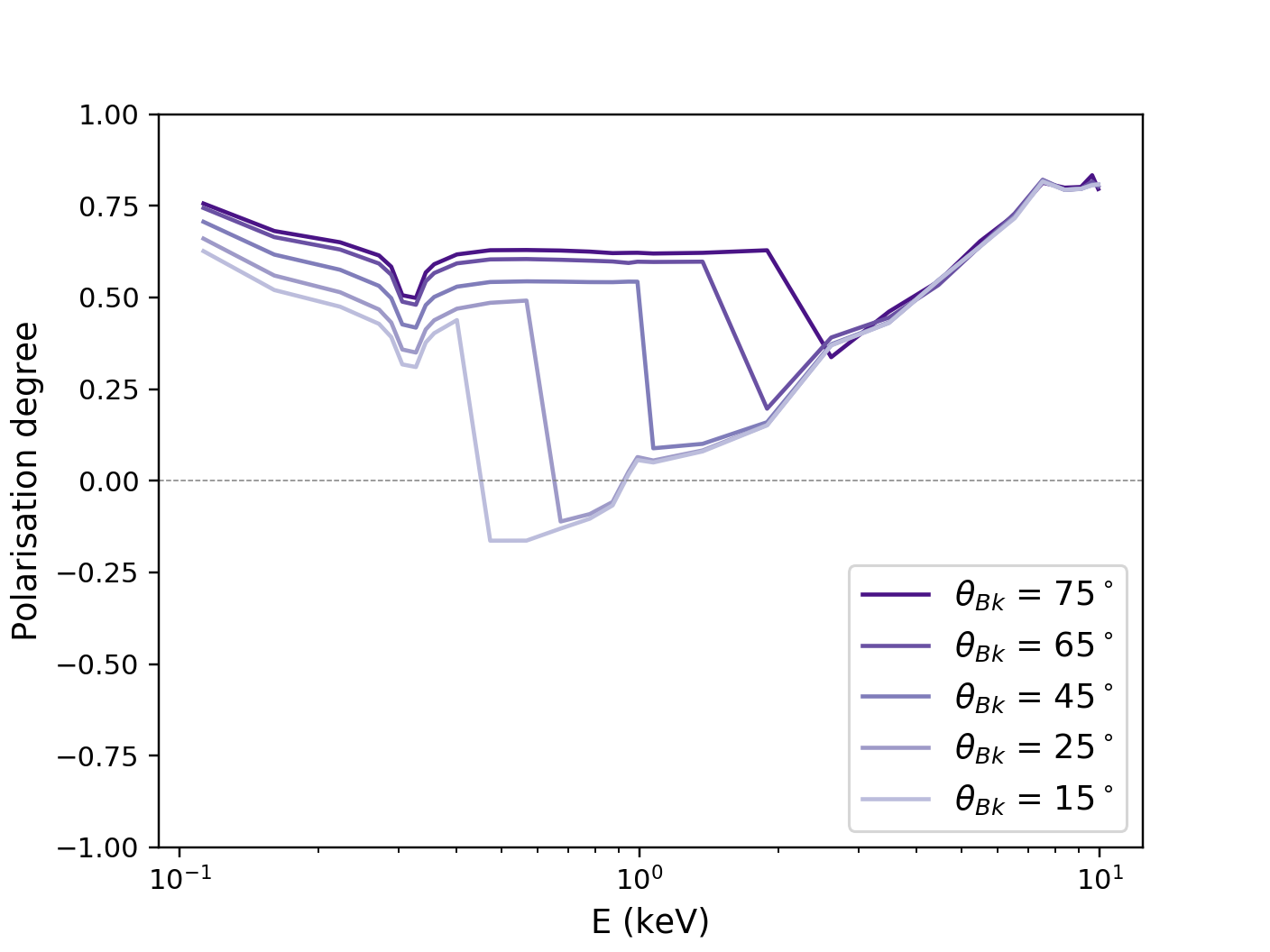}
    \caption{Same as in Figure~\ref{fig:Prob}, for different values of $\theta_\mathrm{Bk}$ in Equation (\ref{Eq:Ead}). Here we used an effective temperature $T=10^7\,\mathrm K$, magnetic field strength $B \approx 5\times10^{13}\, \mathrm G$ and inclination $\theta_\mathrm{B}=0^\circ$.}
    \label{fig:Theta}
\end{figure}

Figure \ref{fig:Prob} shows the result of changing the value of the probability criteria on the polarisation degree of the radiation emerging from a partial mode conversion model with an effective temperature $T = 10^7$ K, a magnetic field strength $B\approx5\times10^{13}$ G, field inclination $\theta_\mathrm{B}=0^\circ$. It can be seen that increasing the required value of $P_\mathrm{con}$ results in a spectrum with a polarisation that is more highly dominated in the X-mode. Additionally, the dependence of the polarisation degree on the assumed value of $\theta_\mathrm{Bk}$ in the evaluation of $E_\mathrm{ad}$ (Equation~\ref{Eq:Ead}) can be seen in Figure \ref{fig:Theta}. As expected from Equations (\ref{Eq:Ead}) and (\ref{Econ}), with an increase in $\theta_\mathrm{Bk}$, the energy at which the probability criteria is met also increases. From both of these trends it can be seen that, while the behavioural trends would remain the same, a different choice in angle and probability criteria would vary the earlier model 
results. As it can be seen, for $B=5 \times 10^{13}$~G, the location of the sudden drop of the polarisation degree at $E_\mathrm{con}$ moves between $\sim 500$~eV and $\sim 3$~keV while $\theta_\mathrm{Bk}$ is increased, leading to the expectation that, in a more realistic simulation, the drop may appear as a smooth decrease over a similar energy range.  Significantly, however, this effect is not sufficient to cause a switch in the dominant mode in the X-ray band for larger values of the magnetic field: for a magnetic field strength $\gtrsim 10^{14}\, \mathrm G$, the polarisation degree of the emission above $1\, \mathrm{keV}$ will remain X-mode dominated for all the probability criteria values.

Finally, for completeness, we also examined the case of a model atmosphere computed including both plasma and vacuum contributions but performing the radiative transfer calculation assuming no mode conversion. This scenario was produced by setting the energy at which the probability criterion is met to be greater than $10\, \mathrm{keV}$ and therefore beyond the upper bound in energy used in our simulations, resulting in an atmosphere containing both plasma and vacuum in which no mode conversion is occurring. Results are shown in Figure \ref{fig:Prob} (curve labelled ``Standard+vac``). As it can be seen, the inclusion of vacuum corrections to the pure-plasma atmosphere results in a reduction of the polarisation degree in the range of $\sim10$--$20\%$ across the entire energy range for a model with $B_\mathrm p\approx5\times10^{13}\,\mathrm{G}$ and $T = 10^7\,\mathrm{K}$. This can be understood because the several peaks caused by the vacuum contribution in the X-mode opacity makes the atmosphere less transparent to this mode. 

\section{Application to Sources} \label{sec:application}

The calculations presented thus far concerned the polarisation properties of an atmospheric layer at constant $T$ and $\boldsymbol B$ as evaluated by a local observer, at rest on the star surface. In this section we investigate the polarisation pattern at infinity for models with no vacuum corrections or including vacuum corrections and assuming complete or partial mode conversion considering different emission geometries from the neutron star surface. 
We assume a core-centred dipole magnetic field, including general relativistic corrections \cite[e.g.][]{page_surface_1996}, with a polar value $B_\mathrm p = 10^{14}\, \mathrm G$ . In strong magnetic fields the suppression of heat transport by electron conduction across the field lines produces a meridional temperature distribution $T_\mathrm{dip}=T_\mathrm p\vert\cos\theta_\mathrm B\vert^{1/2}$ , where $T_\mathrm p$ is the 
temperature at the pole \cite[]{greenstein_pulselike_1983}. The previous, simple expression is in good agreement with the more accurate result by \cite{potekhin_neutron_2015} for magnetic colatitudes $\theta\lesssim 80^\circ$ while it provides unrealistically small values of $T$ near the magnetic equator. To account for this we introduce a minimum temperature $T_\mathrm c$ and define the surface temperature as $T
= \max (T_\mathrm{dip},T_\mathrm c )$; $T_\mathrm{p}=10^{7}\, \mathrm K$ and $T_\mathrm{c}=7\times 10^{6}\, \mathrm K$ were actually used in our calculations.  The star mass and radius were taken to be $M=1.4\, M_\odot$ and $R= 10\, \mathrm{km}$, corresponding to a  surface gravity $g=2.4\times10^{14}\, \mathrm{cm/s^2}$. When computing models with partial mode conversion, in this section we used $\tan \theta_\mathrm{Bk} =1$ in the evaluation of 
$E_\mathrm{ad}$ and $P_\mathrm{con}=0.1$. 

We divided the surface into six annular patches in latitude in such a way that the magnetic field inclination in each patch is $\theta_\mathrm{B} = (0^\circ, 10^\circ, 30^\circ, 50^\circ, 70^\circ, 89^\circ)$; the patches are centred at magnetic colatitudes $(0^\circ, 19.4^\circ, 49.1^\circ, 67.2^\circ, 79.7^\circ, 89.5^\circ)$. For each patch we then computed the atmospheric model for the corresponding values of $B$ and $T$, assumed to be constant within the patch.

In order to obtain the spectral and polarisation properties at infinity, we follow the ray tracing technique detailed in \citet [][see also \citealt{taverna_polarization_2015, gonzalez_caniulef_polarized_2016}]{zane_unveiling_2006} which computes the monochromatic, phase-dependent flux of the three Stokes parameters  by summing together the contributions from all the surface patches which are in view at a certain rotational phase. The observed flux, polarisation degree and angle depend on the star geometry through the angles $\chi$ and $\xi$ that the spin axis makes with the line-of-sight and the magnetic axis, respectively.  

\begin{figure*}
  \centering
  \subfigure[]{\includegraphics[width=\columnwidth]{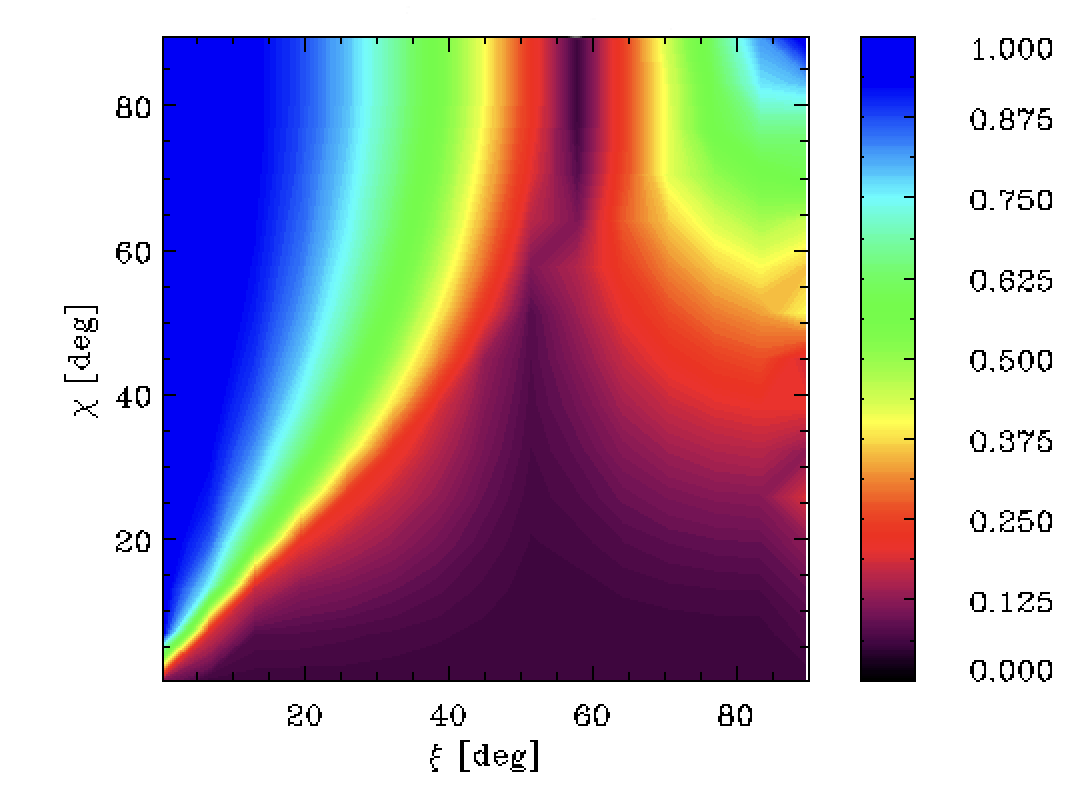}}
  \subfigure[]{\includegraphics[width=\columnwidth]{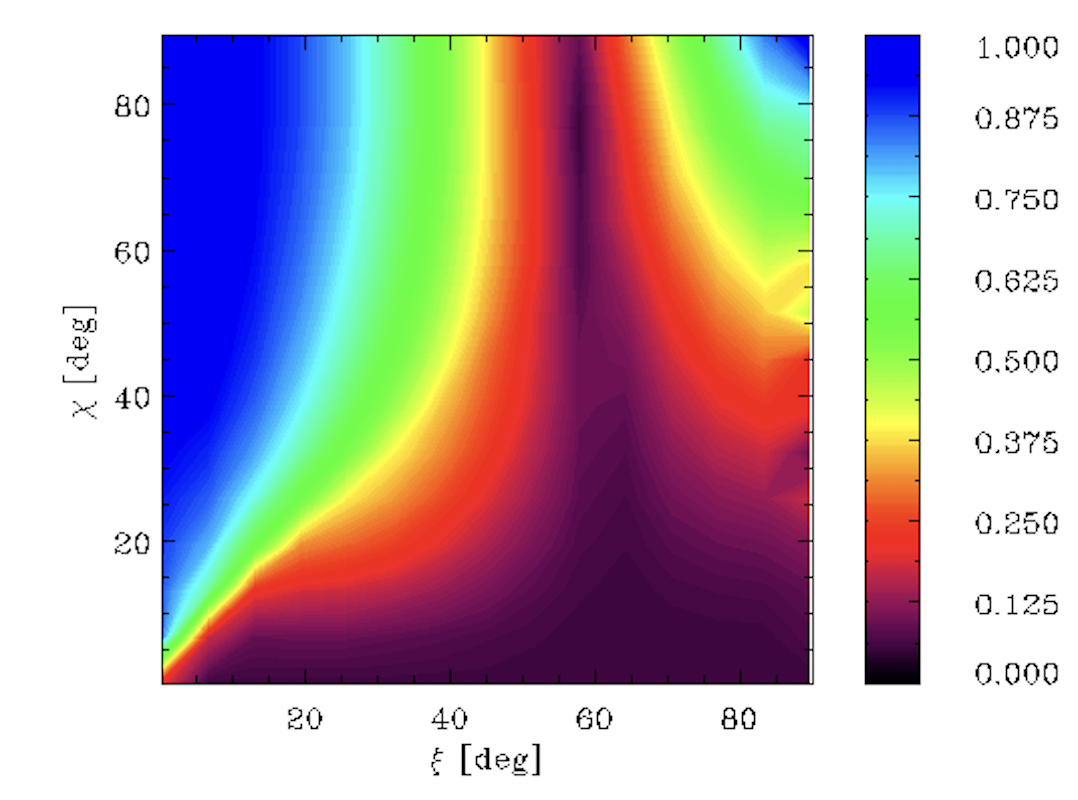}}  
  \subfigure[]{\includegraphics[width=\columnwidth]{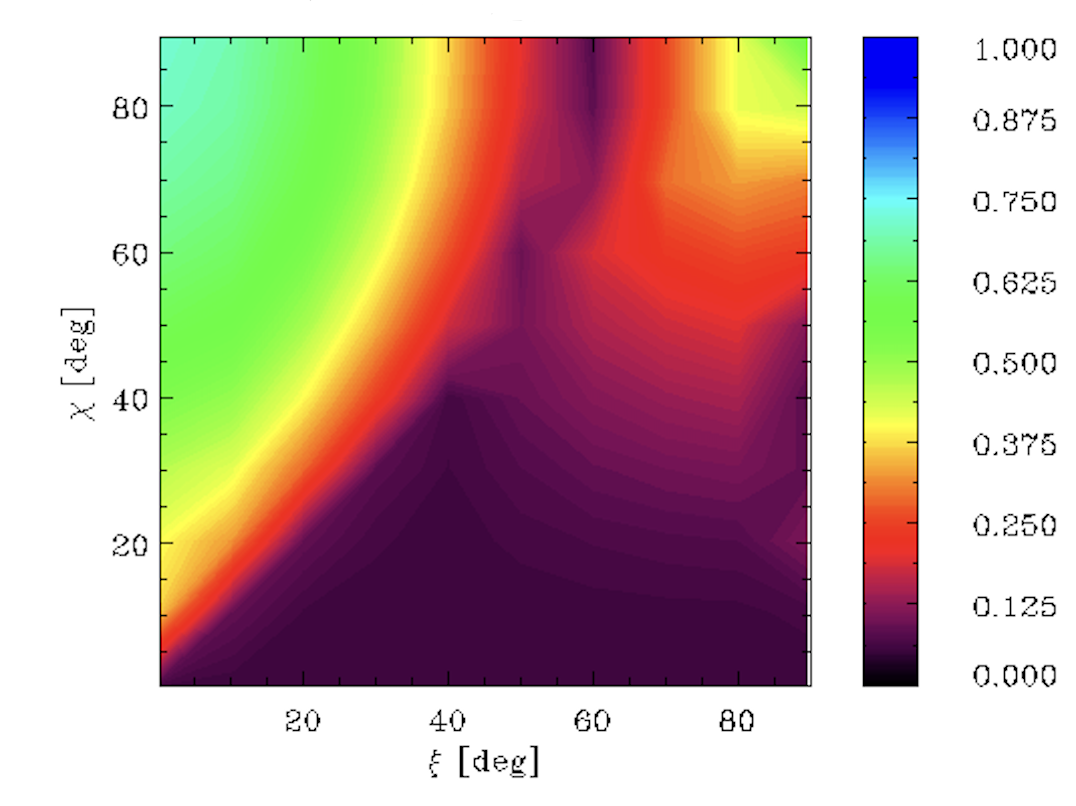}}
  \subfigure[]{\includegraphics[width=\columnwidth]{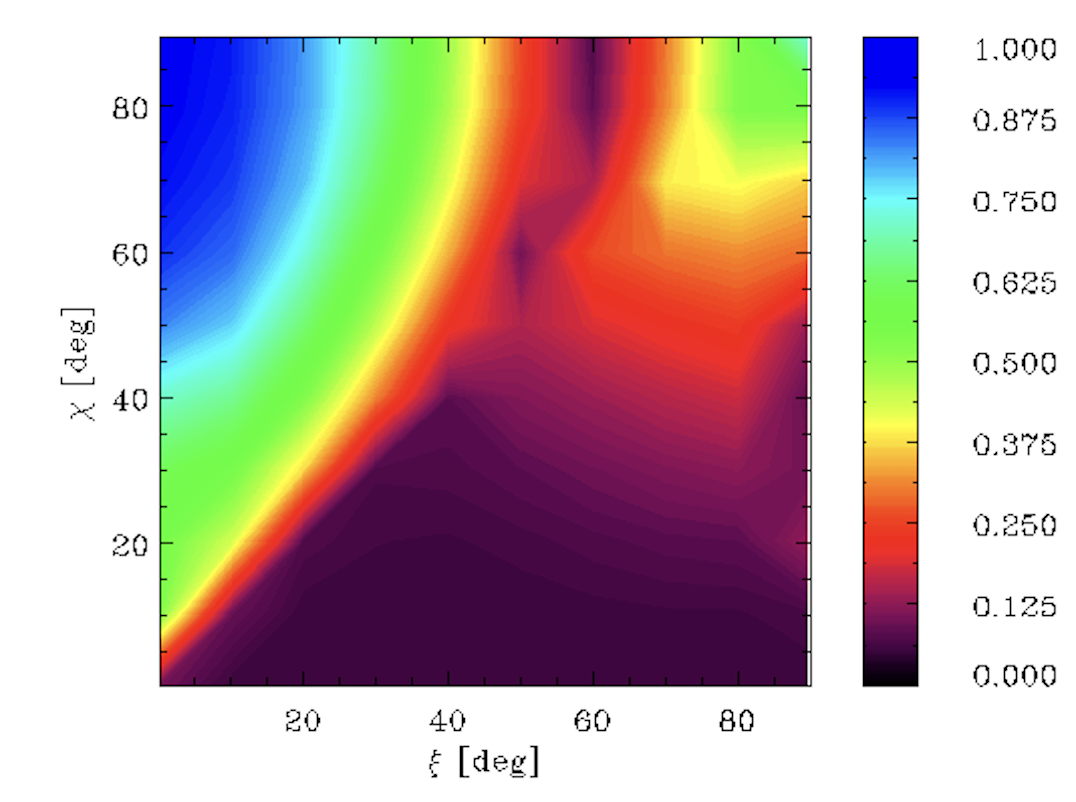}}
  \subfigure[]{\includegraphics[width=\columnwidth]{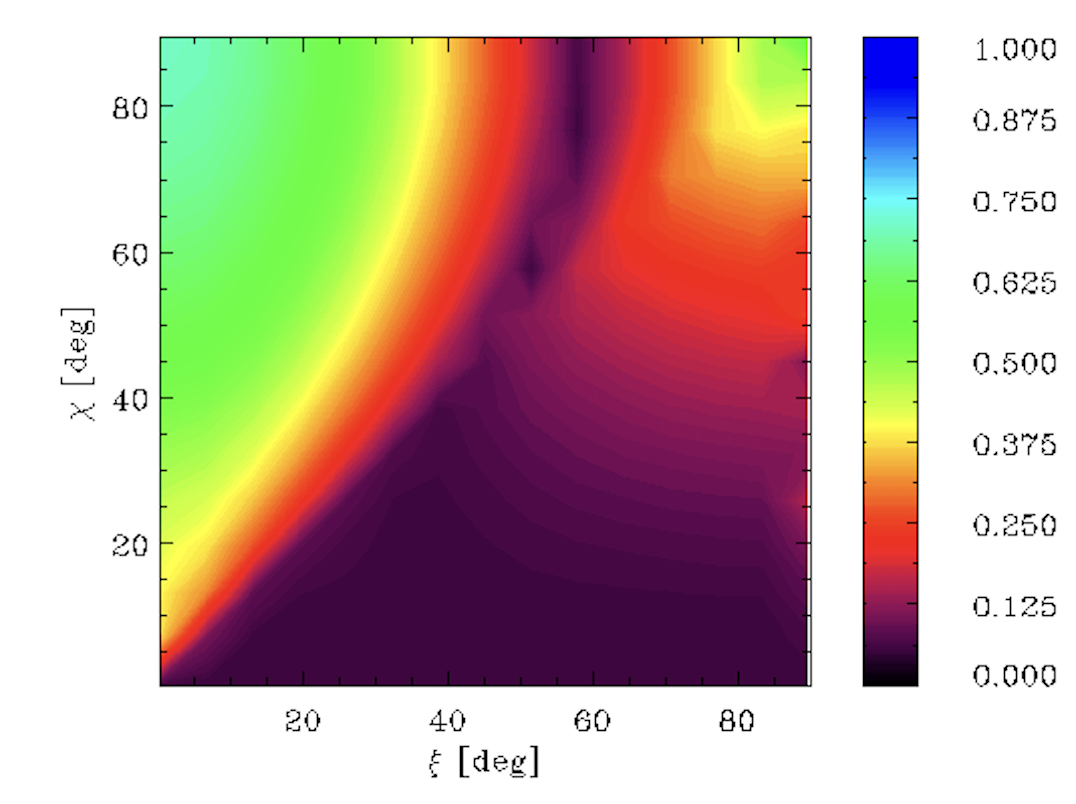}}
  \subfigure[]{\includegraphics[width=\columnwidth]{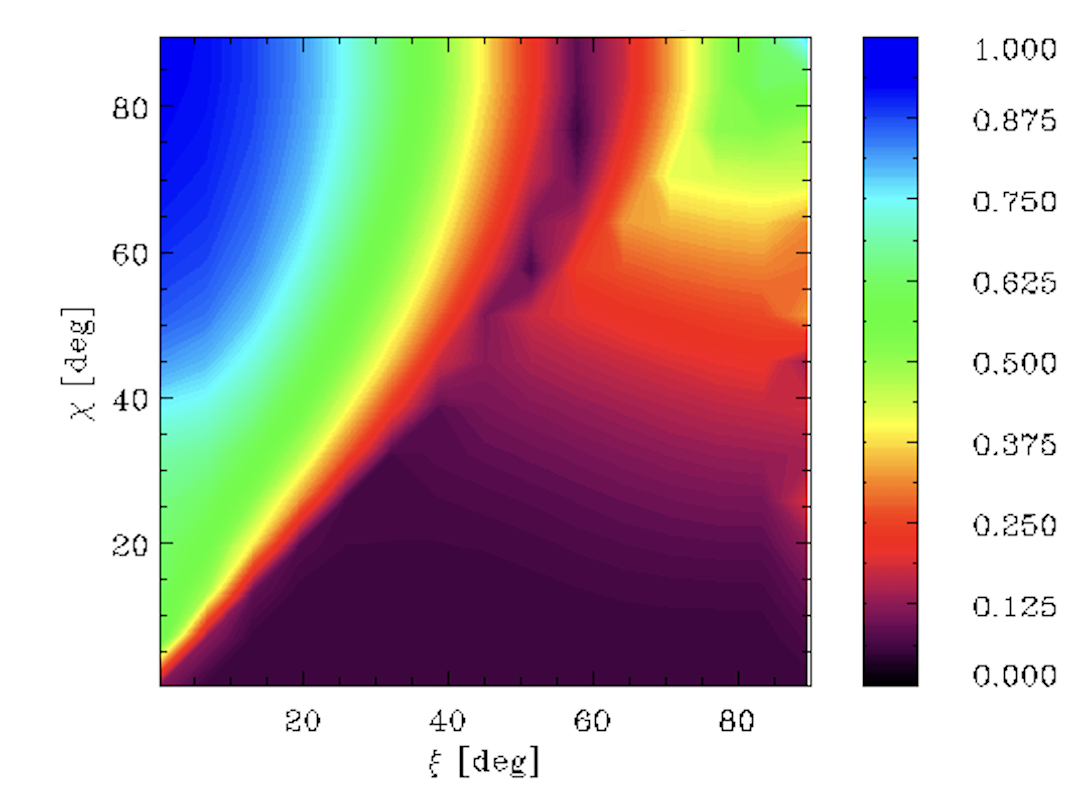}}
  
  \caption{Contour plots of the phase-averaged polarisation degree, as measured by a distant observer, for emission from the whole surface (case A) in the $2$--$4\,\mathrm{keV}$ (panels a, c and e) and $6$--$8\,\mathrm{keV}$ (second column; panels b, d and f) energy bands, for different values of the two angles $\chi$ and $\xi$. Panels a and b refer to a pure plasma atmosphere, panels c and d to an atmosphere with vacuum contribution and  complete mode conversion,  and panels e and f to the case of partial mode conversion. All models use a dipole magnetic field $B=10^{14}\,\mathrm G$ and $T_\mathrm p=10^7\, \mathrm K$.}
  \label{fig:IXPEbandB14}
\end{figure*}

We first consider the case of a magnetar 
emitting from the entire star surface (case A hereafter) and make reference to 
the IXPE working energy band \cite[$2$--$8\, \mathrm{keV}$,][]{weisskopf_imaging_2022}. Figure \ref{fig:IXPEbandB14} shows the phase-averaged polarisation degree at infinity for models without vacuum corrections, with vacuum corrections and complete-mode conversion and with vacuum corrections and partial-mode conversion, as a function of the two angles $\chi$ and $\xi$. 
Emission from a standard plasma atmosphere in which there are no vacuum contributions exhbits a high polarisation degree across the entire $2$--$8\, \mathrm{keV}$ range, increasing slightly from $2$--$4$ to $6$--$8 \, \mathrm{keV}$ (see panels a and b of Figure \ref{fig:IXPEbandB14}). Differently, the models with vacuum contributions and both complete (Figure \ref{fig:IXPEbandB14} c and d) and partial mode conversion (Figure \ref{fig:IXPEbandB14} e and f) show a marked increase in polarisation from a maximum of $\sim 60\%$ at $2$--$4\,\mathrm{keV}$  to $\gtrsim 90\%$ at $6$--$8\,\mathrm{keV}$.

Motivated by the emission geometry suggested by \cite{taverna_polarized_2022} and \cite{zane_strong_2023} to explain the polarisation of 4U 0142+61 and 1RXS J1708, respectively, we also simulated, for the same model parameters, the emission  from a polar cap (called case B hereafter) and an equatorial belt (case C hereafter), both with a semi-aperture of $5^\circ$. In all three scenarios (plasma, complete and partial mode conversion), case B remains highly polarised across the considered energy range, only showing a slight increase of the polarisation with energy. By contrast, case C has very similar properties to case A. In all of these simulations (cases A, B and C) the polarisation angle remains constant across the entire energy range, and hence, there is no switch in dominant polarisation mode.

\begin{figure*}
    \centering
    \subfigure[$0.1$--$0.5$ keV]{\includegraphics[width=0.3\linewidth]{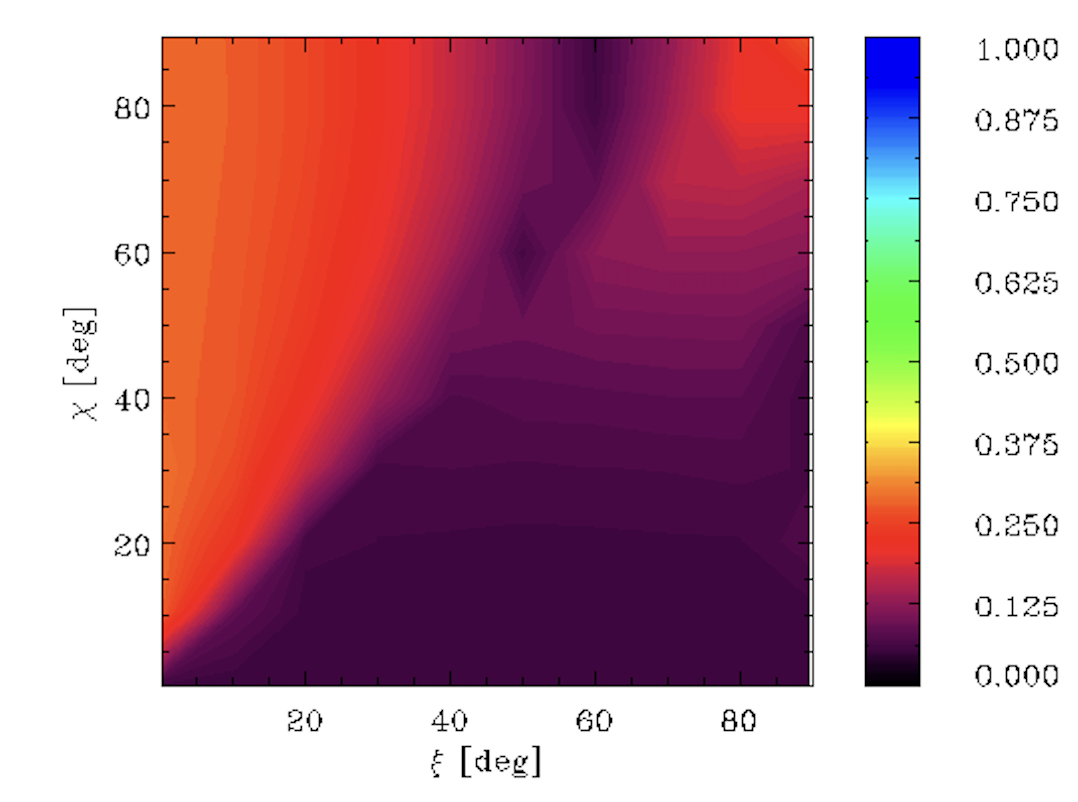}}
    \subfigure[$0.5$--$1$ keV]{\includegraphics[width=0.3\linewidth]{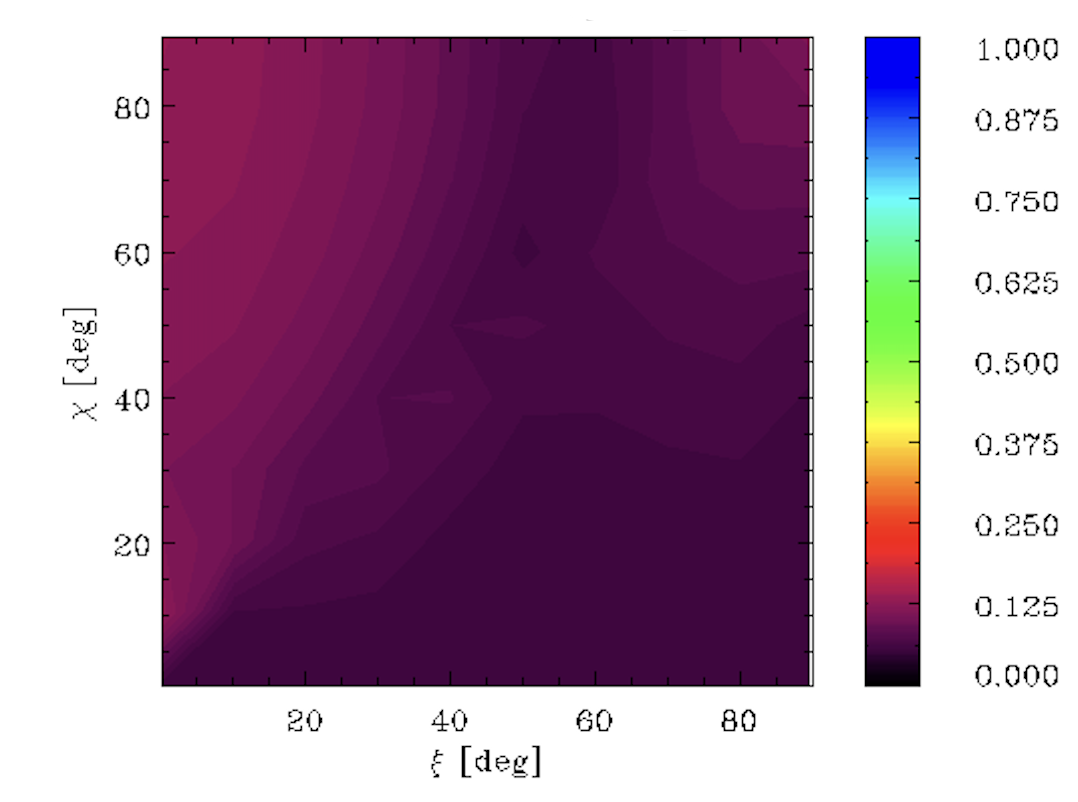}}
    \subfigure[$1$--$2$ keV]{\includegraphics[width=0.3\linewidth]{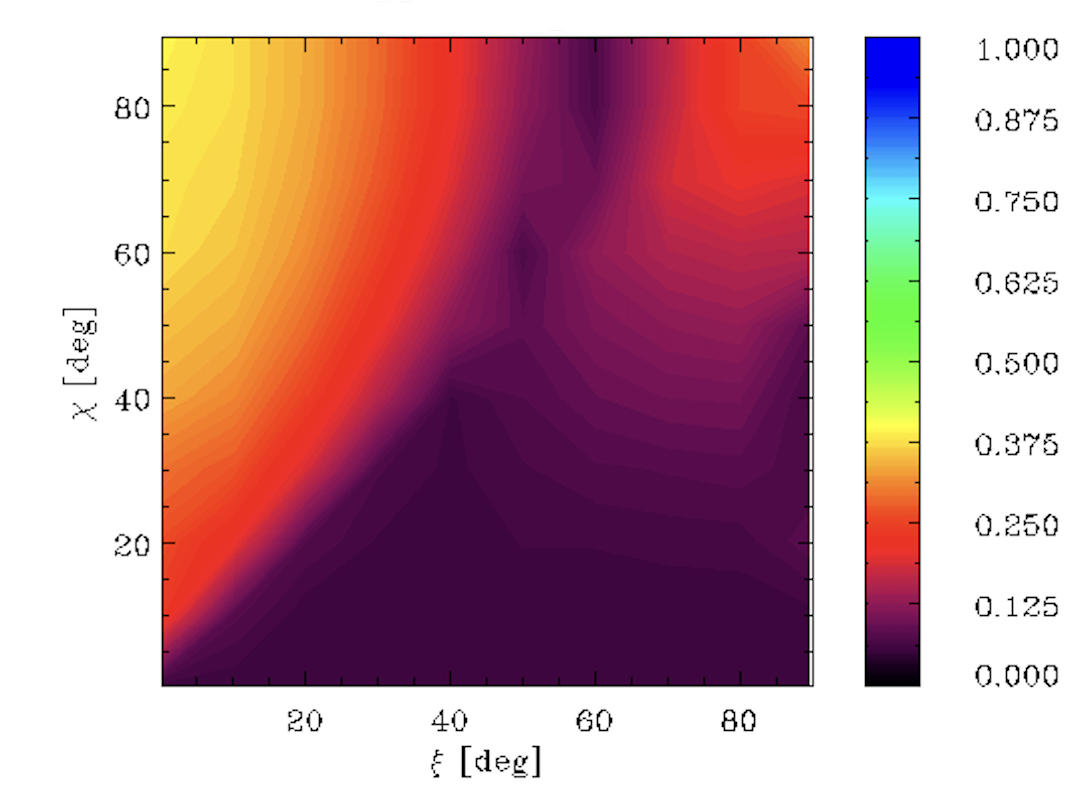}}
    \caption{Same as in Figure \ref{fig:IXPEbandB14} for a model with emission from an equatorial belt of semi-aperture $5^\circ$
    (model C), and complete mode conversion in the energy ranges  $0.1$--$0.5$ (panel a), $0.5$--$1$ (panel b) and $1$--$2\, \mathrm{keV}$ (panel c).}
    \label{fig:B14Belt}
\end{figure*}

\begin{figure*}
    \centering
    \subfigure[$0.1$--$0.5$ keV]{\includegraphics[width=0.3\linewidth]{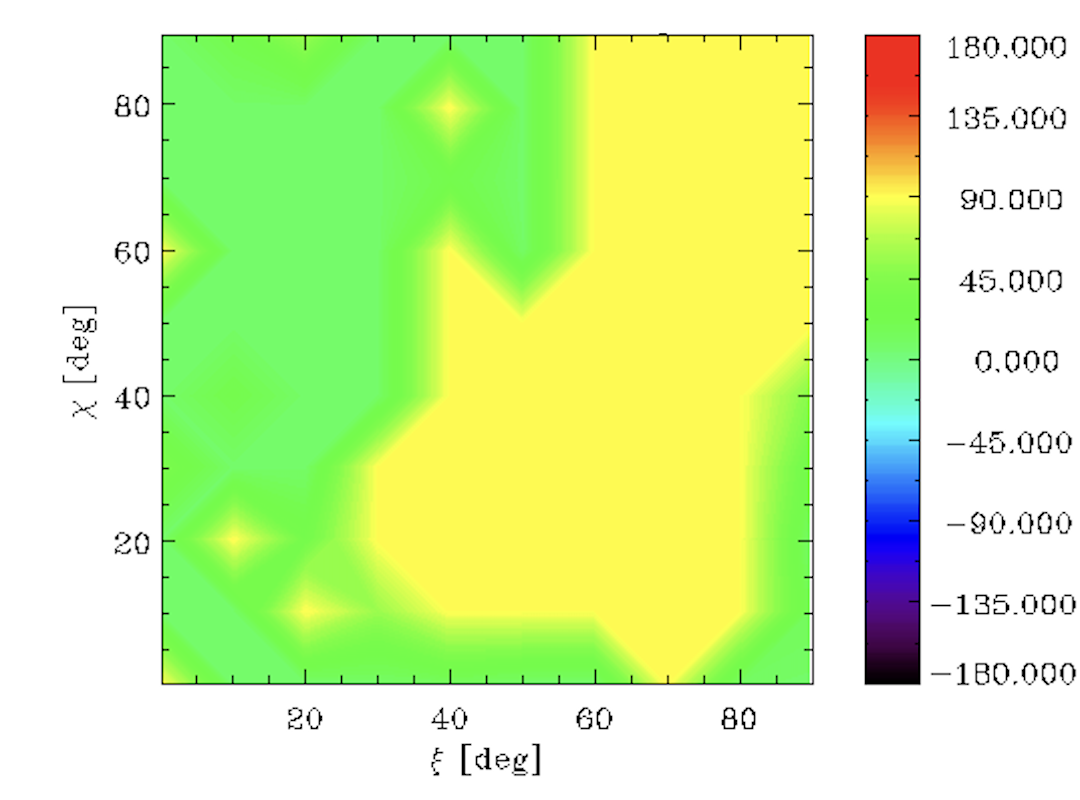}}
    \subfigure[$1$--$2$ keV]{\includegraphics[width=0.3\linewidth]{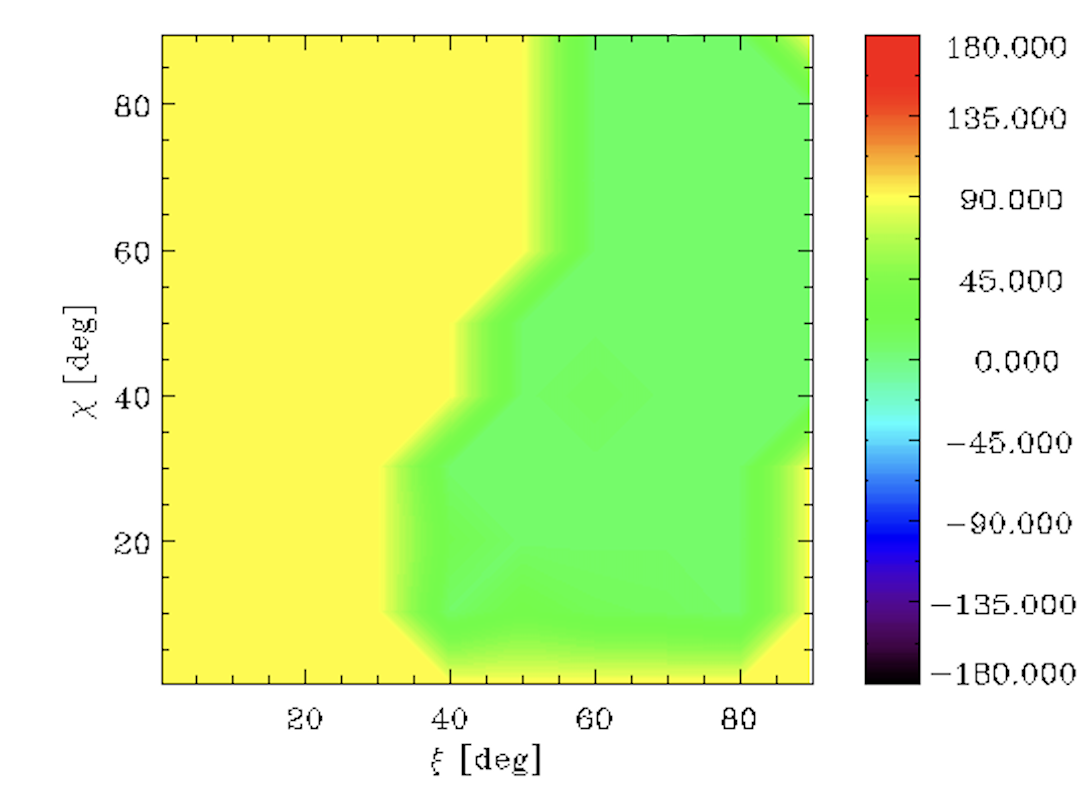}}
    \subfigure[Difference]{\includegraphics[width=0.3\linewidth]{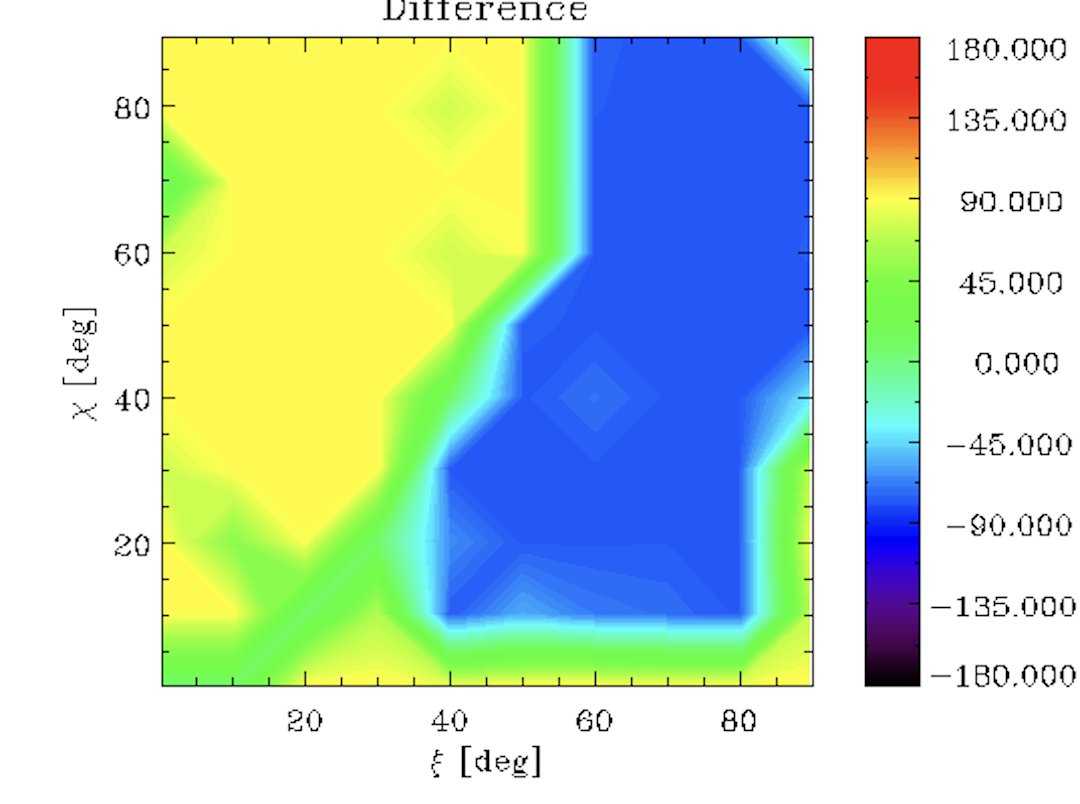}}
    \caption{Same as in Figure \ref{fig:B14Belt} but for the phase-averaged polarisation angle in the energy bands $0.1$--$0.5$ (panel a) and $1$--$2\, \mathrm{keV}$ (panel b); panel (c) shows the difference between the polarisation angles in the two energy ranges.}
    \label{fig:B14BeltAng}
\end{figure*}

However, at energies lower than those detected by IXPE the picture is different. Figure \ref{fig:B14Belt} shows the phase-averaged polarisation degree for complete mode conversion (case C discussed above) in the lower energy ranges $0.1$--$0.5$, $0.5$--$1$ and $1$--$2\, \mathrm{keV}$. Unlike in the higher energy range, the polarisation degree does not monotonically increase with energy. Instead, the maximum polarisation degree is $\sim 25\%$ in the $0.1$--$0.5\, \mathrm{keV}$ range, drops to $\approx 0\%$ around $0.5$--$1$ keV and then rises to $\sim 40\%$ in the $1$--$2\, \mathrm{keV}$ band. At the same time, the polarisation angle shows a $90^\circ$ change between $0.1$--$0.5$ and $1$--$2\, \mathrm{keV}$ (Figure \ref{fig:B14BeltAng}), clearly indicating a switch in the dominant polarisation mode. Due to the higher magnetic field at the pole, with respect to the equator, in the same energy range case B shows no swing in polarisation angle and therefore no change in the dominant polarisation mode.
\begin{figure}
    \centering
    \includegraphics[width=\columnwidth]{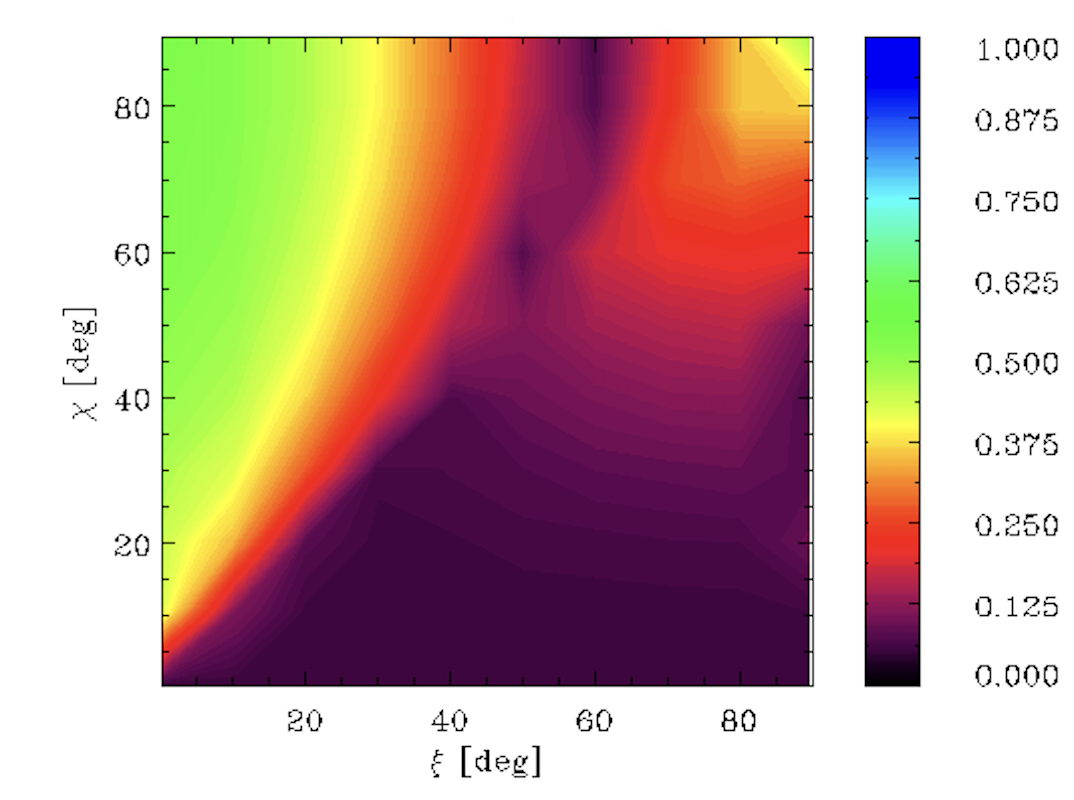}
    \caption{Phase-averaged polarisation degree in the $0.1$--$1\,\mathrm{keV}$ range of emission from an equatorial belt of radius $5^\circ$ 
    (case C),
    of a magnetar atmosphere with partial mode conversion. Here $B_\mathrm p=10^{14}\, \mathrm{G}$.}
\label{fig:PartB14(0.1-1)}
\end{figure}
By comparison, case C with partial mode conversion 
has similar polarisation properties to those with full mode conversion in the $1$--$2$ keV range, but a higher maximum polarisation of $\sim 60\%$ in the  $0.1$--$1\, \mathrm{keV}$ range (Figure \ref{fig:PartB14(0.1-1)}), since these energies fall below the probability threshold for mode conversion. The standard plasma atmosphere model assuming no mode conversion remains highly polarised across the lower energy range. 

For all surface geometries (case A, B and C), both the plasma atmosphere and partial mode conversion models do not show a change in polarisation angle and the emission is therefore dominated by the same polarisation mode across the entire energy range ($0.1$--$8\, \mathrm{keV}$).

\begin{figure*}
    \centering
    \subfigure[$0.1$--$1$ keV]{\includegraphics[width=0.24\linewidth]{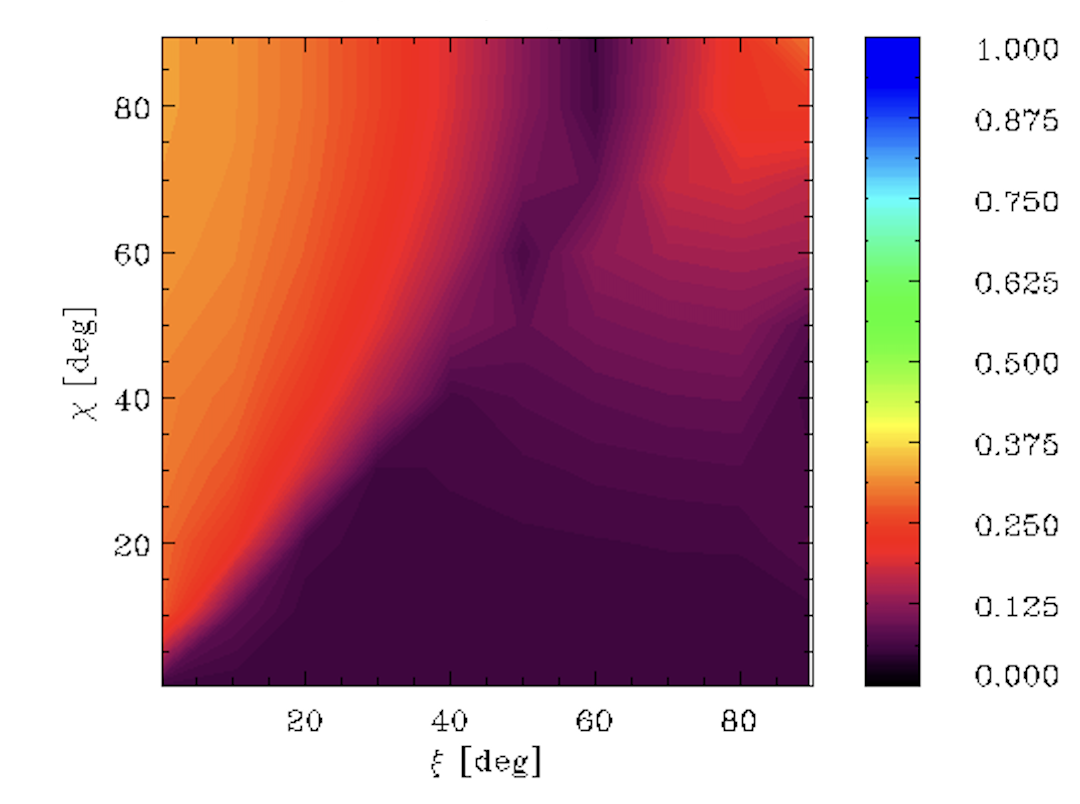}}
    \subfigure[$1$--$2$ keV]{\includegraphics[width=0.24\linewidth]{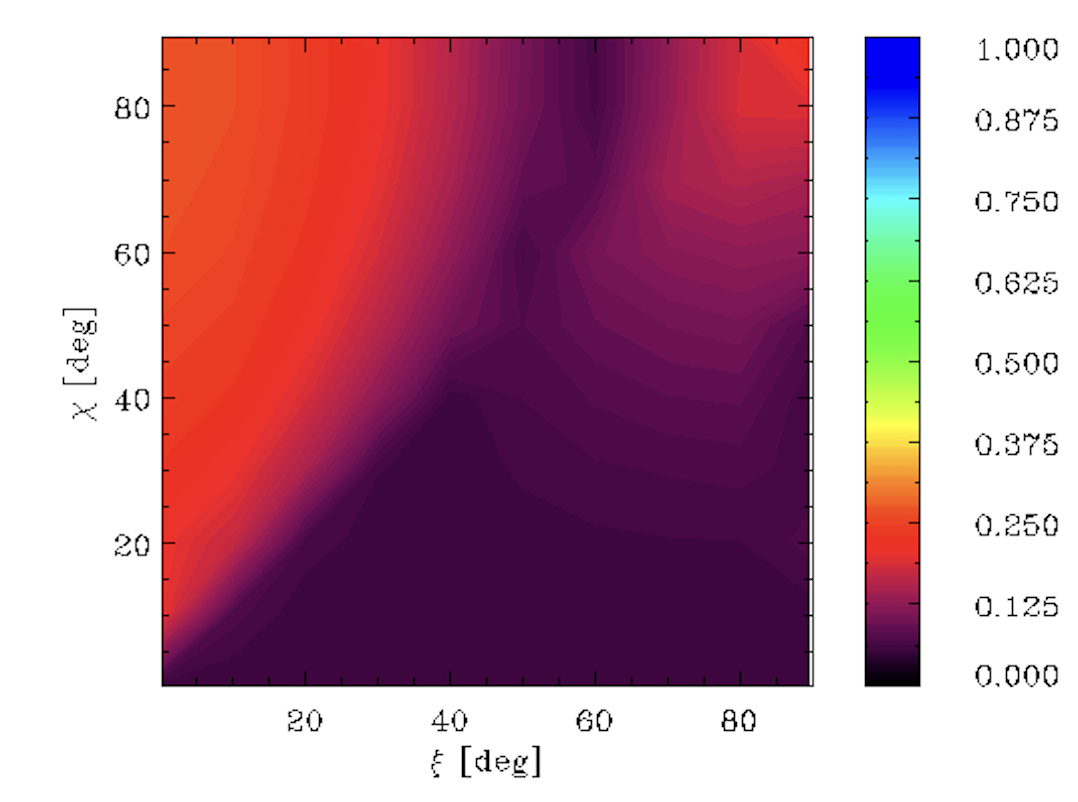}}
    \subfigure[$2$--$4$ keV]{\includegraphics[width=0.24\linewidth]{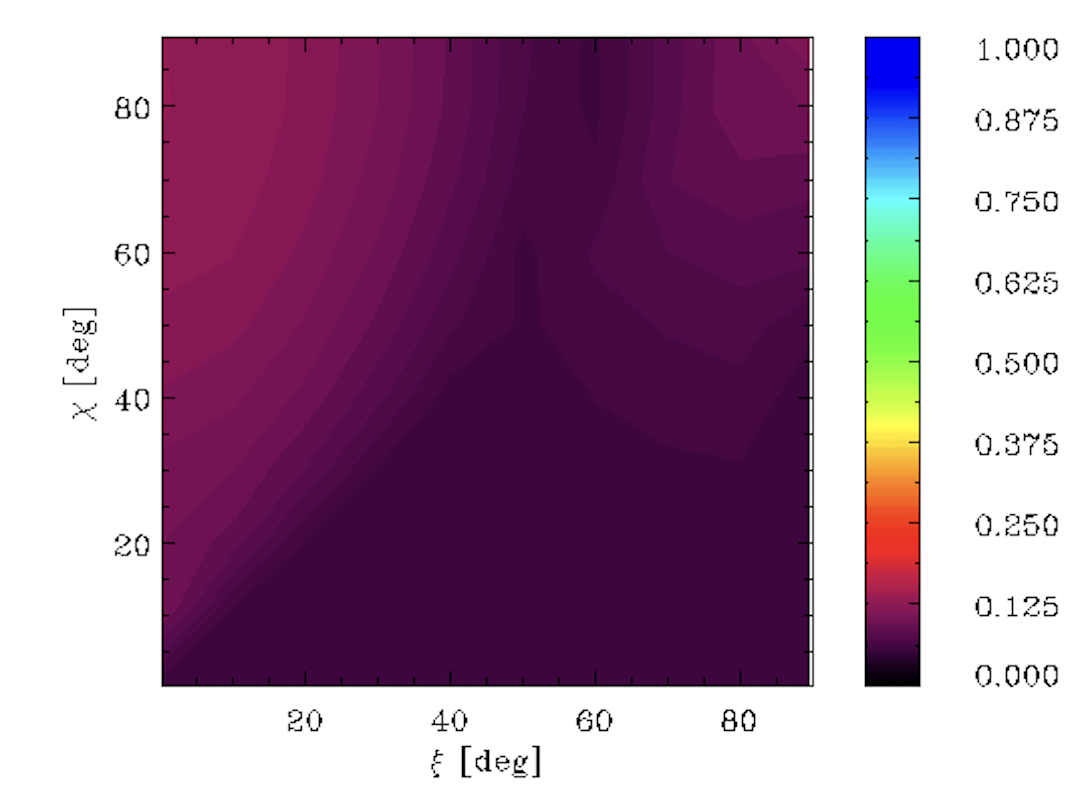}}
    \subfigure[$6$--$8$ keV]{\includegraphics[width=0.24\linewidth]{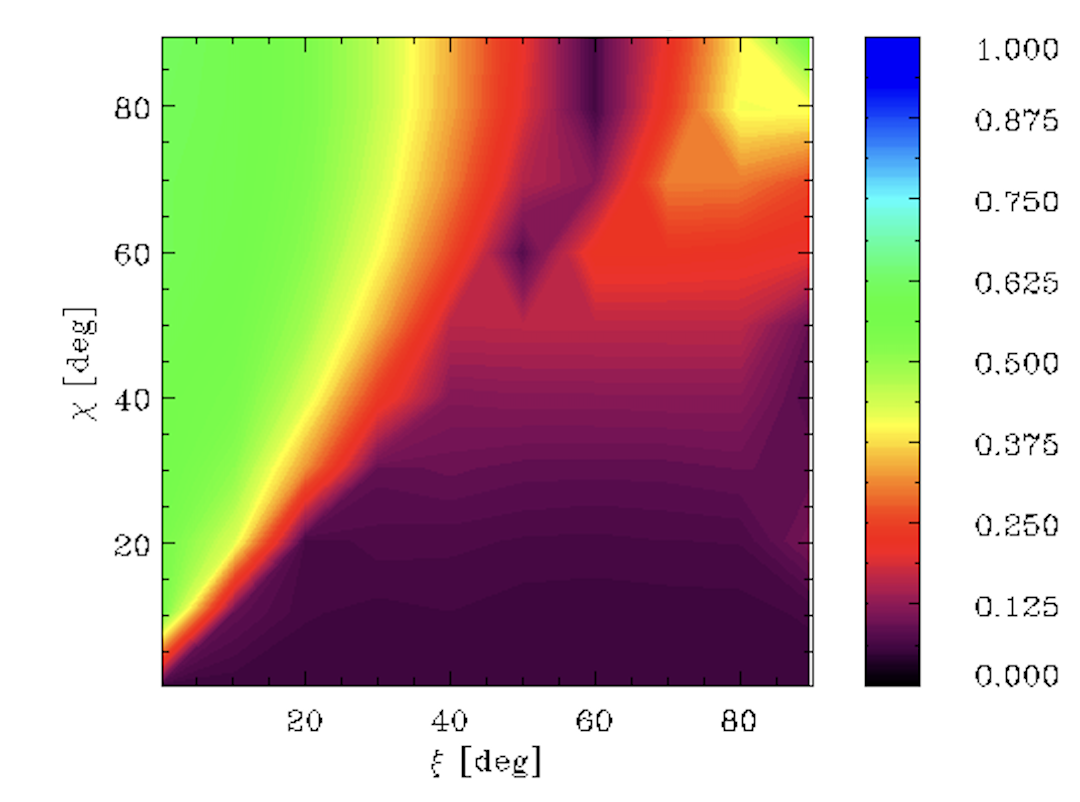}}
    \caption{Phase-averaged polarisation degree of emission from an equatorial belt (case C) with a radius of $5^\circ$ of a magnetar with a dipole magnetic field $B_p=5\times10^{13}$ G and $T_p=10^7$ K and complete mode conversion occurring in the atmosphere.}
    \label{fig:B13.7BeltFull}
\end{figure*}

We additionally produced simulations with a lower dipole magnetic field ($B_\mathrm p = 5\times10^{13}\, \mathrm{G}$). As expected, emission remains highly polarised ($\gtrsim 90\%$) for all the three emission geometries (cases A, B and C) for a standard plasma atmosphere with no vacuum corrections. The models with full mode conversion show a change in the dominant mode across the energy range. However, in agreement with our previous findings (see Figure \ref{fig:Bfields}), the switch occurs at higher energies, for a given surface geometry, when compared with the higher field model ($B_\mathrm p = 10^{14}\mathrm G$). For case C with full mode conversion, the polarisation degree is $\approx25\%$ in one mode (the O-mode) at $0.1$--$1\,\mathrm{keV}$, monotonically decreases  until $2$--$4\mathrm{keV}$, where the it drops to $\sim 0\%$ and the polarisation angle rotates by $90^\circ$. At higher energies the
polarisation is dominated by the X-mode and increases  to $\sim 60\%$ at $6$--$8\,\mathrm{keV}$ (Figure \ref{fig:B13.7BeltFull}).

\begin{figure*}
    \centering
    \subfigure[]{\includegraphics[width=0.3\linewidth]{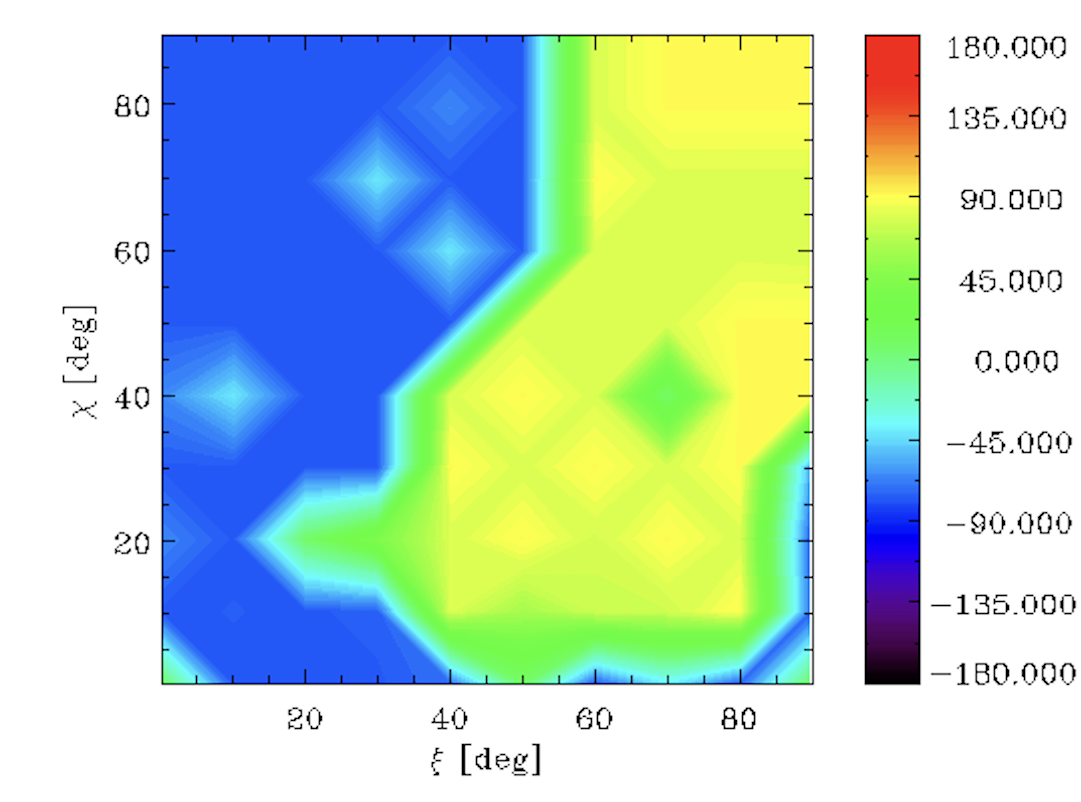}}
    \subfigure[]{\includegraphics[width=0.3\linewidth]{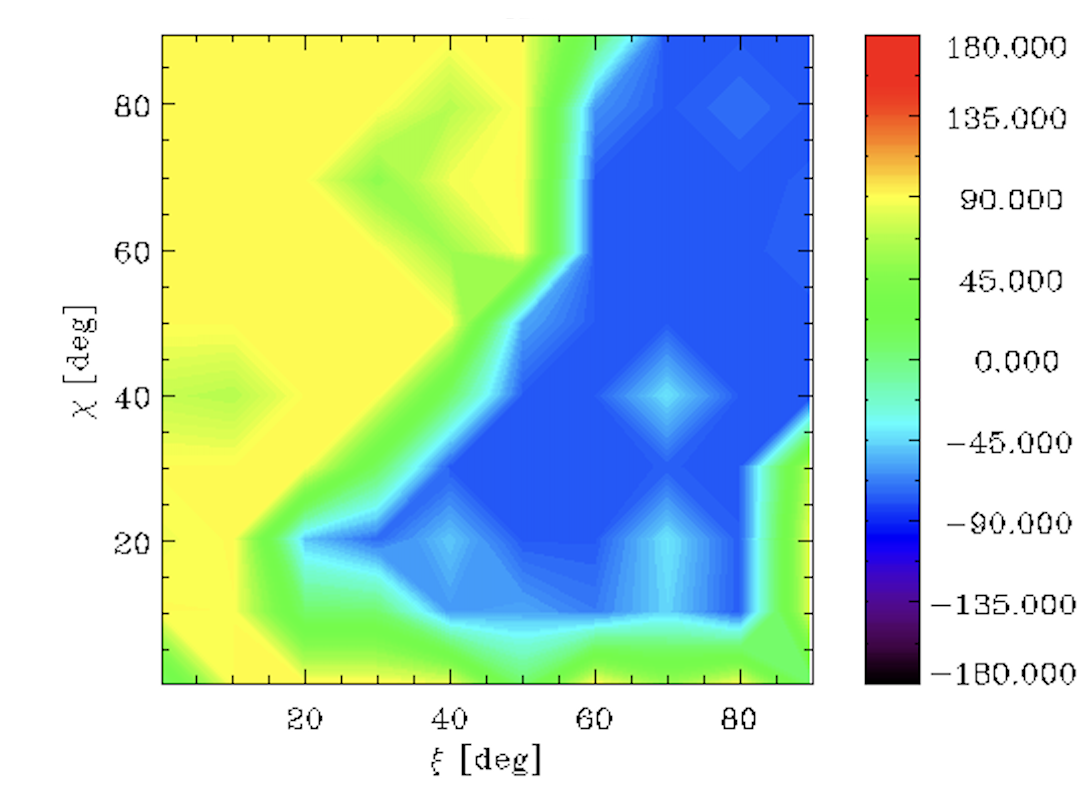}}
    \subfigure[]{\includegraphics[width=0.3\linewidth]{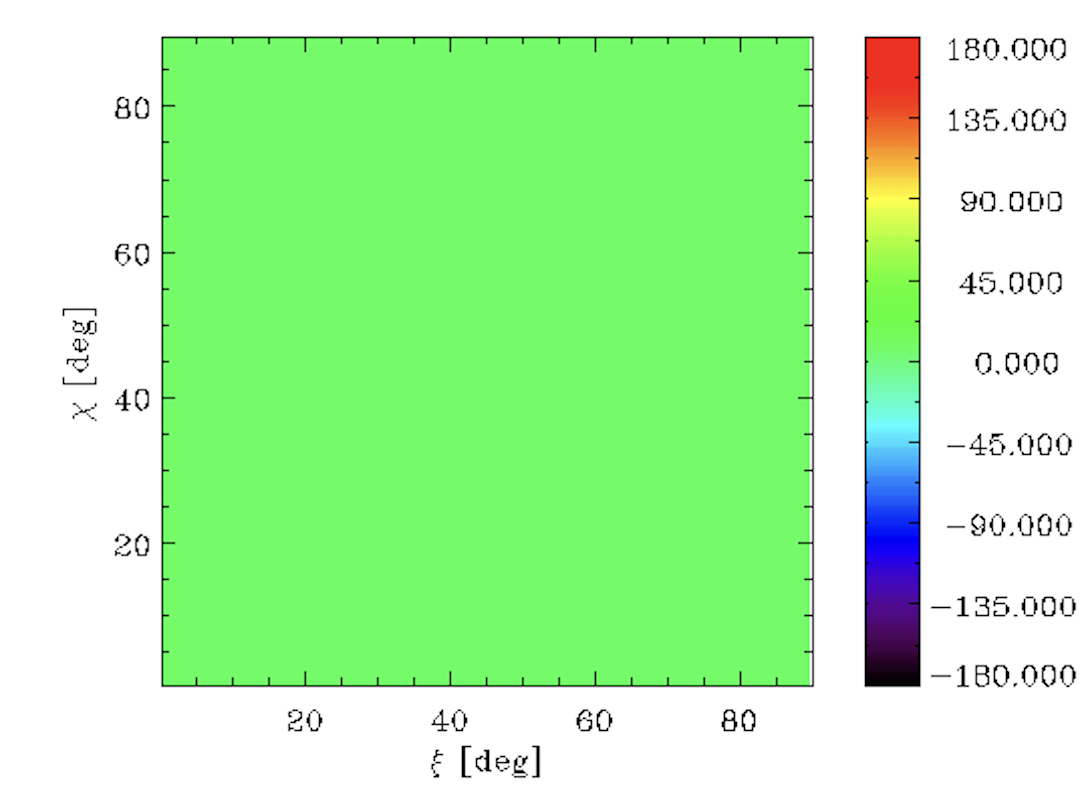}}
    \caption{Difference plots for the phase-averaged polarisation angle for equatorial belt emission (case C) of an atmosphere with partial mode conversion and $B=5\times10^{13}$ G, for the energy bands $0.1$--$1$ and $1$--$2\,\mathrm{keV}$  (a), $1$--$2$ and $6$--$8\,\mathrm{keV}$  (b), and $0.1$--$1$ and $6$--$8\,\mathrm{keV}$ (c).}
    \label{fig:B13.7PartDiff}
\end{figure*}

The model with partial mode conversion, for the same emission geometries, shows a polarisation degree which is similar to the full mode conversion case above but is higher below $\sim 1\,\mathrm{keV}$, i.e. where the mode conversion probability threshold is met, much like what is seen for the $B_\mathrm p = 10^{14}\,\mathrm{G}$ atmospheres. In cases A and B the polarisation angle does not change among the different energy intervals across the entire energy range. Interestingly, however, with partial mode conversion, two $90^\circ$ changes in polarisation angle can be seen for case C. The difference between the polarisation angle at $0.1$--$1$ and $1$--$2\,\mathrm{keV}$, $1$--$2$ and $6$--$8\,\mathrm{keV}$ and $0.1$--$1$ and $6$--$8\,\mathrm{keV}$  is shown in Figure \ref{fig:B13.7PartDiff}. The polarisation angle, and therefore, the dominant polarisation mode (in this case the X-mode) is the same in the $0.1$--$1$ and $6$--$8\,\mathrm{keV}$ range, but the angle is rotated by $90^\circ$ at $1$--$2$ keV and the emission in this energy band is therefore dominated by the other polarisation mode (O-mode).

\begin{table*}
 \caption{Main polarisation features for the three atmosphere models  and  for the different emission geometries: whole surface  (case A), polar cap (case B) and  equatorial belt (case C), the latter two with a semi-aperture of $5^\circ$.}
    \label{tab:ResultsApp}
    \centering
    \begin{tabular}{|c|c|c|c|}
        \hline
        \hline
          & {Standard Plasma} & {Partial Mode Conversion} & {Complete Mode Conversion} \\
        \hline
        \hline
        $B_\mathrm p=10^{14}\, \mathrm{G}$ & & & \\
        \hline
        A & High pol. degree & High pol. degree - $E < 1\, \mathrm{keV}$ & Pol. degree increase with $E$ \\
         & No pol. angle change & Sharp drop in pol. degree - $E \sim 1\, \mathrm{keV}$ & Pol. angle $90^\circ$ rotation between: \\
         & & No pol. angle change & $E = 0.1$--$0.5$ and $0.5$--$1 \mathrm{keV}$ \\
        \hline
        B & High pol. degree & High pol. degree & Pol. degree $\simeq 0$ - $E = 0.1$--$0.5 \, \mathrm{keV}$ \\
         & No pol. angle change & No pol. angle change & Pol. degree increase with $E$ \\
         & & & No pol. angle change \\
        \hline
        C & High pol. degree & High pol. degree - $\mathrm{E} < 1\, \mathrm{keV}$ & Pol. degree drop to $\simeq 0$ - $E = 0.5$--$1 \, \mathrm{keV}$ \\
         & No pol. angle change & Sharp drop in pol. degree - $E \sim 1\, \mathrm{keV}$ & Pol. angle $90^\circ$ rotation - $E = 0.5$--$1\, \mathrm{keV}$\\
         & & No pol. angle change & \\
        \hline
        \hline
        $B_\mathrm p=5\times10^{13}\, \mathrm{G}$ & & & \\
        \hline
        A & High pol. degree & Pol. degree increase with $E$ & Pol. degree reduces to $\simeq 0$ - $E = 1$--$2 \, \mathrm{keV}$ \\
         & No pol. angle change & No pol. angle change & Pol. angle $90^\circ$ rotation - $E = 1$--$2\, \mathrm{keV}$ \\
        \hline
        B & High pol. degree & Pol. degree increase with $E$ & Pol. degree reduces to $\simeq 0$ - $E = 0.5$--$1 \, \mathrm{keV}$ \\
         & No pol. angle change & No pol. angle change & Pol. angle $90^\circ$ rotation - $E = 0.5$--$1\, \mathrm{keV}$\\
        \hline
        C & High pol. degree & High pol. degree - $E < 1\, \mathrm{keV}$ & Pol. degree reduces to $\simeq 0$ - $E = 2$--$4 \, \mathrm{keV}$ \\
         & No pol. angle change & Sharp drop in pol. degree - $E \sim 1\, \mathrm{keV}$ & Pol. angle $90^\circ$ rotation - $E = 2$--$4\, \mathrm{keV}$ \\
         & & Pol. degree reduces to $\simeq 0$ - $E = 2$--$4 \, \mathrm{keV}$ & \\
         & & Two pol. angle $90^\circ$ rotations between: & \\
         & & $0.1$--$1\, \mathrm{keV}$ and $1$-$2\, \mathrm{keV}$ & \\
         & & $1$--$2\, \mathrm{keV}$ and $4$--$8\, \mathrm{keV}$ & \\
        \hline
        \hline
    \end{tabular}
\end{table*}

Table \ref{tab:ResultsApp} summarises the main features for all three atmosphere models and all three emission geometries. Significantly, for the partial mode conversion models we computed, the only scenario in which there is a dominant mode switch is case C for a field $B_\mathrm p=5\times10^{13}\, \mathrm{G}$ where there are two $90^\circ$ polarisation angle rotations. In every other scenario the dominant mode remains the same across the entire energy band. On the other hand, complete mode conversion results in a single $90^\circ$ rotation in the polarisation angle (i.e. a switch in dominant polarisation mode) in almost every model scenario. However, this dominant mode switch only occurs in the IXPE energy range again for case C and $B_\mathrm p=5\times10^{13}\, \mathrm{G}$. In all the other cases, the polarisation angle rotation occurs at energies below those detected by IXPE.

\section{Discussion and Conclusions} \label{sec:discussion}

In this work, we have presented numerical models of magnetar atmospheres either with or without vacuum corrections and considering both complete and partial adiabatic mode conversion. We investigated the spectral and polarisation properties at the surface of the star and at infinity. Our results indicate that mode conversion significantly impacts the polarisation of emission from highly magnetised neutron star atmospheres.

We found that standard plasma atmospheres in which there are no vacuum contributions produce highly polarised emission ($\sim 100\%$ at the star surface and $\gtrsim 80\%$ at infinity) dominated by photons in the X-mode. The polarisation degree increases slightly with energy over the $0.1$--$10\,\mathrm{keV}$  range and there is no change in polarisation angle. The inclusion of vacuum corrections  does affect the polarisation, reducing the polarisation degree.

Our results also show that complete mode conversion results in emission which is dominated by the O-mode at lower energies, becomes less polarised with increasing energy, until it switches to X-dominated and increases in polarisation again. The switch occurs at $2$--$3\, \mathrm{keV}$ for a field strength of $B_\mathrm p=3\times10^{13}$G but moves to lower energies as the magnetic field strength increases. At infinity a lower overall polarisation degree with respect to the standard plasma atmosphere models can be seen at higher energies and a $90^\circ$ rotation in the polarisation angle within the $0.1$--$8\,\mathrm{keV}$ range is present for some emission geometries with field strengths low enough.

Partial mode conversion results in polarisation which is similar to the full mode conversion case above the probability threshold value. Below this value, the emission is always X dominated, although the degree of polarisation is reduced with respect to the standard, no mode conversion, models. Additionally, for magnetic fields $\lesssim5\times10^{13}\,\mathrm{G}$, two switches in the dominant mode (from X-mode to O-mode and then back to X-mode) are present as the polarisation angle rotates by $90^\circ$. For higher magnetic fields, no switch in the dominant mode occurs and the emission remains X-mode dominated across the entire energy range.

We also found that none of the models discussed in this paper is sufficient to explain the emission detected by IXPE from 4U 0142+61 or 1RXS J1708 in the entire IXPE energy band with a single component, unless the values of the dipole magnetic fields inferred from the timing are overestimated. In the IXPE range, neither partial nor complete mode conversion results in a $90^\circ$ rotation in polarisation angle for a magnetar with magnetic field strength $\gtrapprox 10^{14}$ G. Additionally, although both partial and complete mode conversion result in an increase in polarisation degree with energy across the $2$--$8$ keV range, 
this increase ($\approx 30\%$) is not large enough to explain the polarisation degree of the emission from J1708 at high energies, at least for the parameter values we explored.

It is our prediction that, in the IXPE energy band, mode conversion may result in more sizeable effects  in neutron stars with lower magnetic fields such as the X-ray Dim Isolated Neutron Stars \cite[XDINSs, see e.g.][]{turolla_isolated_2009}. Additionally, this phenomenon will likely be very relevant in connection with the Rocket Experiment Demonstration of a Soft X-ray Polarimeter (REDSoX) which is expected to detect X-ray polarisation in the $0.2$--$0.8\, \mathrm{keV}$ range \citep{marshall_globe_2021}.

There is ample room for improvement in our work. Most notably, the treatment of partial mode conversion relies on a fixed probability threshold due to the complications in including angles in our probability function which would affect  evaluation of the scattering integrals. The result of this is a step function, with a sharp change in polarisation at the threshold energy at which every photon meets the criteria for mode conversion to occur. In reality, the emission would have a much smoother evolution; we expect the polarisation degree would gradually reduce, with less O-mode photons present at energies below $E_\mathrm{ad}$, where the probability of mode conversion is close to unity, after which the polarisation spectrum would be in agreement with our results. We also caveat that we assumed a fully ionized, pure H atmosphere in our calculations. The main purpose of this work is in fact to isolate and analyse the impact of different assumptions about mode conversion at the vacuum resonance on the expected polarisation signal. To this end, we assumed a simple model for the plasma composition (pure, ionized H). This scenario may be less realistic for cool, highly magnetised isolated neutron stars, in which case more realistic calculations should account for the effect of partial ionization \citep{ho_iii_2003, potekhin_electromagnetic_2004}. This calculation is beyond the purpose of this paper and will be a matter of future work.  

\section*{Acknowledgements}

RK acknowledges The Science and Technology Facilities Council (STFC) for funding through a PhD studentship. RK additionally acknowledges University of Padova for hosting a visit to the Department of Physics and Astronomy where much of this work was undertaken. She would also like to thank Dr Nabil Brice for some helpful discussions. The work of RTu and RTa is partially supported by the grant PRIN-2022LWPEXW of the Italian MUR. The authors thank the referee A. Potekhin for his careful reading of the manuscript and helpful comments.

\section*{Data Availability}

The simulated data produced throughout this investigation are available on request.



\bibliographystyle{mnras}
\bibliography{references}

\begin{thebibliography}{}
\makeatletter
\relax
\def\mn@urlcharsother{\let\do\@makeother \do\$\do\&\do\#\do\^\do\_\do\%\do\~}
\def\mn@doi{\begingroup\mn@urlcharsother \@ifnextchar [ {\mn@doi@} {\mn@doi@[]}}
\def\mn@doi@[#1]#2{\def\@tempa{#1}\ifx\@tempa\@empty \href {http://dx.doi.org/#2} {doi:#2}\else \href {http://dx.doi.org/#2} {#1}\fi \endgroup}
\def\mn@eprint#1#2{\mn@eprint@#1:#2::\@nil}
\def\mn@eprint@arXiv#1{\href {http://arxiv.org/abs/#1} {{\tt arXiv:#1}}}
\def\mn@eprint@dblp#1{\href {http://dblp.uni-trier.de/rec/bibtex/#1.xml} {dblp:#1}}
\def\mn@eprint@#1:#2:#3:#4\@nil{\def\@tempa {#1}\def\@tempb {#2}\def\@tempc {#3}\ifx \@tempc \@empty \let \@tempc \@tempb \let \@tempb \@tempa \fi \ifx \@tempb \@empty \def\@tempb {arXiv}\fi \@ifundefined {mn@eprint@\@tempb}{\@tempb:\@tempc}{\expandafter \expandafter \csname mn@eprint@\@tempb\endcsname \expandafter{\@tempc}}}

\bibitem[\protect\citeauthoryear{Adler}{Adler}{1971}]{adler_photon_1971}
Adler S.~L.,  1971, \mn@doi [Annals of Physics] {10.1016/0003-4916(71)90154-0}, 67, 599

\bibitem[\protect\citeauthoryear{Caiazzo, González-Caniulef, Heyl  \& Fernández}{Caiazzo et~al.}{2022}]{caiazzo_probing_2022}
Caiazzo I.,  González-Caniulef D.,  Heyl J.,   Fernández R.,  2022, \mn@doi [Monthly Notices of the Royal Astronomical Society] {10.1093/mnras/stac1571}, 514, 5024

\bibitem[\protect\citeauthoryear{Colpi \& Wasserman}{Colpi \& Wasserman}{2002}]{colpi_formation_2002}
Colpi M.,  Wasserman I.,  2002, \mn@doi [The Astrophysical Journal] {10.1086/344405}, 581, 1271

\bibitem[\protect\citeauthoryear{Ginzburg}{Ginzburg}{1970}]{ginzburg_propagation_1970}
Ginzburg V.~L.,  1970, The propagation of electromagnetic waves in plasmas

\bibitem[\protect\citeauthoryear{Gnedin, Pavlov  \& Shibanov}{Gnedin et~al.}{1978}]{gnedin_effect_1978}
Gnedin Y.~N.,  Pavlov G.~G.,   Shibanov Y.~A.,  1978, Soviet Astronomy Letters, 4, 117

\bibitem[\protect\citeauthoryear{González~Caniulef, Zane, Taverna, Turolla  \& Wu}{González~Caniulef et~al.}{2016}]{gonzalez_caniulef_polarized_2016}
González~Caniulef D.,  Zane S.,  Taverna R.,  Turolla R.,   Wu K.,  2016, \mn@doi [Monthly Notices of the Royal Astronomical Society] {10.1093/mnras/stw804}, 459, 3585

\bibitem[\protect\citeauthoryear{González-Caniulef, Zane, Turolla  \& Wu}{González-Caniulef et~al.}{2019}]{gonzalez-caniulef_atmosphere_2019}
González-Caniulef D.,  Zane S.,  Turolla R.,   Wu K.,  2019, \mn@doi [Monthly Notices of the Royal Astronomical Society] {10.1093/mnras/sty3159}, 483, 599

\bibitem[\protect\citeauthoryear{Greenstein \& Hartke}{Greenstein \& Hartke}{1983}]{greenstein_pulselike_1983}
Greenstein G.,  Hartke G.~J.,  1983, \mn@doi [The Astrophysical Journal] {10.1086/161195}, 271, 283

\bibitem[\protect\citeauthoryear{Hambaryan, Wagner, Schmidt, Hohle  \& Neuhäuser}{Hambaryan et~al.}{2015}]{hambaryan_3xmm_2015}
Hambaryan V.,  Wagner D.,  Schmidt J.,  Hohle M.~M.,   Neuhäuser R.,  2015, \mn@doi [Astronomische Nachrichten] {10.1002/asna.201412189}, 336, 545

\bibitem[\protect\citeauthoryear{Harding \& Lai}{Harding \& Lai}{2006}]{harding_physics_2006}
Harding A.~K.,  Lai D.,  2006, \mn@doi [Reports on Progress in Physics] {10.1088/0034-4885/69/9/R03}, 69, 2631

\bibitem[\protect\citeauthoryear{Heyl \& Hernquist}{Heyl \& Hernquist}{1997}]{heyl_analytic_1997}
Heyl J.~S.,  Hernquist L.,  1997, \mn@doi [Physical Review D] {10.1103/PhysRevD.55.2449}, 55, 2449

\bibitem[\protect\citeauthoryear{Heyl et~al.,}{Heyl et~al.}{2024}]{heyl_detection_2024}
Heyl J.,  et~al., 2024, \mn@doi [Monthly Notices of the Royal Astronomical Society] {10.48550/arXiv.2311.03637}, in press

\bibitem[\protect\citeauthoryear{Ho \& Lai}{Ho \& Lai}{2001}]{ho_atmospheres_2001}
Ho W. C.~G.,  Lai D.,  2001, \mn@doi [Monthly Notices of the Royal Astronomical Society] {10.1046/j.1365-8711.2001.04801.x}, 327, 1081

\bibitem[\protect\citeauthoryear{Ho \& Lai}{Ho \& Lai}{2003}]{ho_ii_2003}
Ho W. C.~G.,  Lai D.,  2003, \mn@doi [Monthly Notices of the Royal Astronomical Society] {10.1046/j.1365-8711.2003.06047.x}, 338, 233

\bibitem[\protect\citeauthoryear{Ho, Lai, Potekhin  \& Chabrier}{Ho et~al.}{2003}]{ho_iii_2003}
Ho W. C.~G.,  Lai D.,  Potekhin A.~Y.,   Chabrier G.,  2003, \mn@doi [The Astrophysical Journal] {10.1086/379507}, 599, 1293

\bibitem[\protect\citeauthoryear{Ho, Chang, Kaplan, Mori, Potekhin  \& van Adelsberg}{Ho et~al.}{2007}]{ho_modeling_2007}
Ho W. C.~G.,  Chang P.,  Kaplan D.~L.,  Mori K.,  Potekhin A.~Y.,   van Adelsberg M.,  2007, \mn@doi [Advances in Space Research] {10.1016/j.asr.2007.04.011}, 40, 1432

\bibitem[\protect\citeauthoryear{Kaspi \& Beloborodov}{Kaspi \& Beloborodov}{2017}]{kaspi_magnetars_2017}
Kaspi V.~M.,  Beloborodov A.~M.,  2017, \mn@doi [Annual Review of Astronomy and Astrophysics] {10.1146/annurev-astro-081915-023329}, 55, 261

\bibitem[\protect\citeauthoryear{Lai}{Lai}{2001}]{lai_matter_2001}
Lai D.,  2001, \mn@doi [Reviews of Modern Physics] {10.1103/RevModPhys.73.629}, 73, 629

\bibitem[\protect\citeauthoryear{Lai}{Lai}{2023}]{lai_ixpe_2023}
Lai D.,  2023, \mn@doi [Proceedings of the National Academy of Sciences] {10.1073/pnas.2216534120}, 120, e2216534120

\bibitem[\protect\citeauthoryear{Lai \& Ho}{Lai \& Ho}{2002}]{lai_resonant_2002}
Lai D.,  Ho W. C.~G.,  2002, \mn@doi [The Astrophysical Journal] {10.1086/338074}, 566, 373

\bibitem[\protect\citeauthoryear{Lai \& Ho}{Lai \& Ho}{2003}]{lai_transfer_2003}
Lai D.,  Ho W. C.~G.,  2003, \mn@doi [The Astrophysical Journal] {10.1086/374334}, 588, 962

\bibitem[\protect\citeauthoryear{Liu, Ng  \& Dodson}{Liu et~al.}{2023}]{liu_radio_2023}
Liu Y.~H.,  Ng C.-Y.,   Dodson R.,  2023, \mn@doi [The Astrophysical Journal] {10.3847/1538-4357/acb20d}, 945, 82

\bibitem[\protect\citeauthoryear{Lloyd}{Lloyd}{2003}]{lloyd_model_2003}
Lloyd D.~A.,  2003, \mn@doi [arXiv/astro-ph/0303561] {10.48550/ARXIV.ASTRO-PH/0303561}

\bibitem[\protect\citeauthoryear{Lucy}{Lucy}{1964}]{lucy_temperature-correction_1964}
Lucy L.~B.,  1964, SAO Special Report, 167, 93

\bibitem[\protect\citeauthoryear{Marshall, Marscher, Schulz, Garner, Heine, Masterson  \& Guenther}{Marshall et~al.}{2021}]{marshall_globe_2021}
Marshall H.,  Marscher A.,  Schulz N.~S.,  Garner A.,  Heine S.,  Masterson R.,   Guenther M.,  2021, in 43rd {COSPAR} {Scientific} {Assembly}. p.~1662, \url {https://ui.adsabs.harvard.edu/abs/2021cosp...43E1662M}

\bibitem[\protect\citeauthoryear{Mihalas}{Mihalas}{1978}]{mihalas_stellar_1978}
Mihalas D.,  1978, Stellar atmospheres

\bibitem[\protect\citeauthoryear{Mori \& Ho}{Mori \& Ho}{2007}]{mori_modelling_2007}
Mori K.,  Ho W. C.~G.,  2007, \mn@doi [Monthly Notices of the Royal Astronomical Society] {10.1111/j.1365-2966.2007.11663.x}, 377, 905

\bibitem[\protect\citeauthoryear{Nobili, Turolla  \& Zane}{Nobili et~al.}{2008}]{nobili_x-ray_2008}
Nobili L.,  Turolla R.,   Zane S.,  2008, \mn@doi [Monthly Notices of the Royal Astronomical Society] {10.1111/j.1365-2966.2008.13627.x}, 389, 989

\bibitem[\protect\citeauthoryear{Ozel}{Ozel}{2001}]{ozel_surface_2001}
Ozel F.,  2001, \mn@doi [The Astrophysical Journal] {10.1086/323851}, 563, 276

\bibitem[\protect\citeauthoryear{Page \& Sarmiento}{Page \& Sarmiento}{1996}]{page_surface_1996}
Page D.,  Sarmiento A.,  1996, \mn@doi [The Astrophysical Journal] {10.1086/178216}, 473, 1067

\bibitem[\protect\citeauthoryear{Pavlov \& Shibanov}{Pavlov \& Shibanov}{1979}]{pavlov_effect_1979}
Pavlov G.~G.,  Shibanov Y.~A.,  1979, Soviet Journal of Experimental and Theoretical Physics, 76(5), 1457

\bibitem[\protect\citeauthoryear{Pavlov, Shibanov, Ventura  \& Zavlin}{Pavlov et~al.}{1994}]{pavlov_model_1994}
Pavlov G.~G.,  Shibanov Y.~A.,  Ventura J.,   Zavlin V.~E.,  1994, Astronomy and Astrophysics, 289, 837

\bibitem[\protect\citeauthoryear{Pavlov, Zavlin, Trümper  \& Neuhäuser}{Pavlov et~al.}{1996}]{pavlov_multiwavelength_1996}
Pavlov G.~G.,  Zavlin V.~E.,  Trümper J.,   Neuhäuser R.,  1996, \mn@doi [The Astrophysical Journal] {10.1086/310355}, 472, L33

\bibitem[\protect\citeauthoryear{Posselt et~al.,}{Posselt et~al.}{2017}]{posselt_gemingas_2017}
Posselt B.,  et~al., 2017, \mn@doi [The Astrophysical Journal] {10.3847/1538-4357/835/1/66}, 835, 66

\bibitem[\protect\citeauthoryear{Potekhin, Lai, Chabrier  \& Ho}{Potekhin et~al.}{2004}]{potekhin_electromagnetic_2004}
Potekhin A.~Y.,  Lai D.,  Chabrier G.,   Ho W. C.~G.,  2004, \mn@doi [The Astrophysical Journal] {10.1086/422679}, 612, 1034

\bibitem[\protect\citeauthoryear{Potekhin, Pons  \& Page}{Potekhin et~al.}{2015}]{potekhin_neutron_2015}
Potekhin A.~Y.,  Pons J.~A.,   Page D.,  2015, \mn@doi [Space Science Reviews] {10.1007/s11214-015-0180-9}, 191, 239

\bibitem[\protect\citeauthoryear{Rajagopal \& Romani}{Rajagopal \& Romani}{1996}]{rajagopal_model_1996}
Rajagopal M.,  Romani R.~W.,  1996, \mn@doi [The Astrophysical Journal] {10.1086/177059}, 461, 327

\bibitem[\protect\citeauthoryear{Rea, Viganò, Israel, Pons  \& Torres}{Rea et~al.}{2014}]{rea_3xmm_2014}
Rea N.,  Viganò D.,  Israel G.~L.,  Pons J.~A.,   Torres D.~F.,  2014, \mn@doi [The Astrophysical Journal] {10.1088/2041-8205/781/1/L17}, 781, L17

\bibitem[\protect\citeauthoryear{Romani}{Romani}{1987}]{romani_model_1987}
Romani R.~W.,  1987, \mn@doi [The Astrophysical Journal] {10.1086/165010}, 313, 718

\bibitem[\protect\citeauthoryear{Shibanov, Zavlin, Pavlov  \& Ventura}{Shibanov et~al.}{1992}]{shibanov_model_1992}
Shibanov I.~A.,  Zavlin V.~E.,  Pavlov G.~G.,   Ventura J.,  1992, Astronomy and Astrophysics, 266, 313

\bibitem[\protect\citeauthoryear{Suleimanov, Pavlov  \& Werner}{Suleimanov et~al.}{2012}]{suleimanov_magnetized_2012}
Suleimanov V.~F.,  Pavlov G.~G.,   Werner K.,  2012, \mn@doi [The Astrophysical Journal] {10.1088/0004-637X/751/1/15}, 751, 15

\bibitem[\protect\citeauthoryear{Taverna, Turolla, Gonzalez~Caniulef, Zane, Muleri  \& Soffitta}{Taverna et~al.}{2015}]{taverna_polarization_2015}
Taverna R.,  Turolla R.,  Gonzalez~Caniulef D.,  Zane S.,  Muleri F.,   Soffitta P.,  2015, \mn@doi [Monthly Notices of the Royal Astronomical Society] {10.1093/mnras/stv2168}, 454, 3254

\bibitem[\protect\citeauthoryear{Taverna, Turolla, Suleimanov, Potekhin  \& Zane}{Taverna et~al.}{2020}]{taverna_x-ray_2020}
Taverna R.,  Turolla R.,  Suleimanov V.,  Potekhin A.~Y.,   Zane S.,  2020, \mn@doi [Monthly Notices of the Royal Astronomical Society] {10.1093/mnras/staa204}, 492, 5057

\bibitem[\protect\citeauthoryear{Taverna et~al.,}{Taverna et~al.}{2022}]{taverna_polarized_2022}
Taverna R.,  et~al., 2022, \mn@doi [Science] {10.1126/science.add0080}, 378, 646

\bibitem[\protect\citeauthoryear{Thompson \& Duncan}{Thompson \& Duncan}{1993}]{thompson_neutron_1993}
Thompson C.,  Duncan R.~C.,  1993, \mn@doi [The Astrophysical Journal] {10.1086/172580}, 408, 194

\bibitem[\protect\citeauthoryear{Thompson, Lyutikov  \& Kulkarni}{Thompson et~al.}{2002}]{thompson_electrodynamics_2002}
Thompson C.,  Lyutikov M.,   Kulkarni S.~R.,  2002, \mn@doi [The Astrophysical Journal] {10.1086/340586}, 574, 332

\bibitem[\protect\citeauthoryear{Turolla}{Turolla}{2009}]{turolla_isolated_2009}
Turolla R.,  2009, Isolated {Neutron} {Stars}: {The} {Challenge} of {Simplicity}.
 Astrophysics and {Space} {Science} {Library} Vol. 357, Springer Berlin Heidelberg

\bibitem[\protect\citeauthoryear{Turolla, Zane  \& Watts}{Turolla et~al.}{2015}]{turolla_magnetars_2015}
Turolla R.,  Zane S.,   Watts A.~L.,  2015, \mn@doi [Reports on Progress in Physics] {10.1088/0034-4885/78/11/116901}, 78, 116901

\bibitem[\protect\citeauthoryear{Turolla et~al.,}{Turolla et~al.}{2023}]{turolla_ixpe_2023}
Turolla R.,  et~al., 2023, \mn@doi [The Astrophysical Journal] {10.3847/1538-4357/aced05}, 954, 88

\bibitem[\protect\citeauthoryear{Unsold}{Unsold}{1955}]{unsold_physik_1955}
Unsold A.,  1955, Physik der {Sternatmospharen}, {MIT} besonderer {Berucksichtigung} der {Sonne}.

\bibitem[\protect\citeauthoryear{Ventura}{Ventura}{1979}]{ventura_scattering_1979}
Ventura J.,  1979, \mn@doi [Physical Review D] {10.1103/PhysRevD.19.1684}, 19, 1684

\bibitem[\protect\citeauthoryear{Weisskopf et~al.,}{Weisskopf et~al.}{2022}]{weisskopf_imaging_2022}
Weisskopf M.~C.,  et~al., 2022, \mn@doi [Journal of Astronomical Telescopes, Instruments, and Systems] {10.1117/1.JATIS.8.2.026002}, 8

\bibitem[\protect\citeauthoryear{Zane \& Turolla}{Zane \& Turolla}{2006}]{zane_unveiling_2006}
Zane S.,  Turolla R.,  2006, \mn@doi [Monthly Notices of the Royal Astronomical Society] {10.1111/j.1365-2966.2005.09784.x}, 366, 727

\bibitem[\protect\citeauthoryear{Zane, Turolla  \& Treves}{Zane et~al.}{2000}]{zane_magnetized_2000}
Zane S.,  Turolla R.,   Treves A.,  2000, \mn@doi [The Astrophysical Journal] {10.1086/309027}, 537, 387

\bibitem[\protect\citeauthoryear{Zane, Turolla, Stella  \& Treves}{Zane et~al.}{2001}]{zane_proton_2001}
Zane S.,  Turolla R.,  Stella L.,   Treves A.,  2001, \mn@doi [The Astrophysical Journal] {10.1086/322360}, 560, 384

\bibitem[\protect\citeauthoryear{Zane et~al.,}{Zane et~al.}{2023}]{zane_strong_2023}
Zane S.,  et~al., 2023, \mn@doi [The Astrophysical Journal Letters] {10.3847/2041-8213/acb703}, 944, L27

\bibitem[\protect\citeauthoryear{van Adelsberg \& Lai}{van Adelsberg \& Lai}{2006}]{van_adelsberg_atmosphere_2006}
van Adelsberg M.,  Lai D.,  2006, \mn@doi [Monthly Notices of the Royal Astronomical Society] {10.1111/j.1365-2966.2006.11098.x}, 373, 1495

\makeatother
\end{thebibliography}



\bsp	
\label{lastpage}
\end{document}